\documentclass[sigconf]{acmart}

\usepackage{subfigure}
\usepackage{amsmath}
\usepackage{subeqnarray}
\usepackage{cases}
\usepackage{verbatim}
\usepackage[misc]{ifsym}

\usepackage{multicol}
\usepackage{multirow}
\usepackage{booktabs}
\usepackage{balance}

\def\BibTeX{{\rm B\kern-.05em{\sc i\kern-.025em b}\kern-.08emT\kern-.1667em\lower.7ex\hbox{E}\kern-.125emX}}
    
\setcopyright{acmcopyright}

\begin{document}
\fancyhead{}

\copyrightyear{2019}
\acmYear{2019}
\acmConference[MM '19]{Proceedings of the 27th ACM International Conference on Multimedia}{October 21--25, 2019}{Nice, France}
\acmBooktitle{Proceedings of the 27th ACM International Conference on Multimedia (MM '19), October 21--25, 2019, Nice, France}
\acmPrice{15.00}
\acmDOI{10.1145/3343031.3350945}
\acmISBN{978-1-4503-6889-6/19/10}
%
\title{DTDN: Dual-task De-raining Network}

%

\author{Zheng Wang}
\affiliation{%
  \institution{Beijing Key Laboratory of Intelligent Information Technology}\institution{Beijing Institute of Technology, Beijing 100081, China}
}
\author{Jianwu Li}
\authornote{Corresponding author: Jianwu Li (ljw@bit.edu.cn)}
\affiliation{%
  \institution{Beijing Key Laboratory of Intelligent Information Technology}\institution{Beijing Institute of Technology, Beijing 100081, China}
}
\author{Ge Song}
\affiliation{%
  \institution{Beijing Key Laboratory of Intelligent Information Technology}\institution{Beijing Institute of Technology, Beijing 100081, China}
}

%
\renewcommand{\shortauthors}{Zheng Wang, et al.}

%
\begin{abstract}
Removing rain streaks from rainy images is necessary for many tasks in computer vision, such as object detection and recognition. It needs to address two mutually exclusive objectives: removing rain streaks and reserving realistic details. Balancing them is critical for de-raining methods. We propose an end-to-end network, called dual-task de-raining network (DTDN), consisting of two sub-networks: generative adversarial network (GAN) and convolutional neural network (CNN), to remove rain streaks via coordinating the two mutually exclusive objectives self-adaptively. DTDN-GAN is mainly used to remove structural rain streaks, and DTDN-CNN is designed to recover details in original images. We also design a training algorithm to train these two sub-networks of DTDN alternatively, which share same weights but use different training sets. We further enrich two existing datasets to approximate the distribution of real rain streaks. Experimental results show that our method outperforms several recent state-of-the-art methods, based on both benchmark testing datasets and real rainy images. The code is on \url{https://github.com/long-username/DTDN-DTDN-Dual-task-De-raining-Network}.
\end{abstract}

%
%


%

%

%
\maketitle

\section{Introduction}

Removing rain streaks from images can benefit many tasks in computer vision, such as object detection and recognition. The de-raining methods can be divided into two categories: video de-raining and single image de-raining. Video de-raining is able to utilize existing redundant temporal information in adjacent frames of images, which is helpful for identifying and removing rain streaks. Removing rain streaks from single images, however, faces difficult challenging owing to the ill-posed nature and less available information. 

Given a rainy image, rain streaks may be mistaken for image details during de-raining, vice versa. Removing rain streaks completely and retaining image details as much as possible are mutually exclusive. It is critical for a de-raining method to balance the two tasks during de-raining. Although some previous methods have gained great steps, they more or less ignore the complexity of balancing the two exclusive tasks. A typical case is the work in ~\cite{yang2017deep} which removes rain streaks clearly, even in the presence of heavy rain, but it damages color and textures. An additional fidelity term, mean square error (MSE), is designed for the loss functions of ~\cite{zhang2017image} and~\cite{zhang2018density}, such that the details can be retained better. However, adding a fidelity term directly into loss functions increases the difficulty of optimization, leading to some residual rain streaks. In this paper, we propose a Dual-Task De-raining Network (DTDN) with a training algorithm for better removing rain streaks and retaining details via coordinating automatically the mutually exclusive tasks. The DTDN contains two sub-networks corresponding to the two mutually exclusive tasks, respectively, and the two sub-networks are trained alternatively on different training sets to increase the possibility of obtaining a well-refined de-raining network.


Moreover, two existing large-scale synthetic datasets~\cite{zhang2018density, fu2017removing} only contain single types of rain streaks, but we observe that diverse complex rain streaks appear simultaneously in real images. Therefore, we add some rainy noises into existing rainy images to enrich data, which can represent real complex rainy conditions better. 

The main contributions include:
\begin{itemize}
\item[$\bullet$] We propose an end-to-end network with two sub-networks and a specific training algorithm to remove rain streaks while retaining original detail information. 
\item[$\bullet$] We introduce a technique of data enrichment for existing datasets, which makes our network perform better in complex rainy conditions.
\item[$\bullet$] Extensive experiments show that the proposed method outperforms previous state-of-the-art methods.

\end{itemize}

\section{Related Works}
\subsection{Video De-raining}
Video based de-raining methods are able to discover the difference between adjacent frames, such that it is relatively easy to remove rain streaks from videos. Garg and Nayar~\cite{garg2004detection,garg2005does,garg2007vision} propose an appearance model using photometric properties and temporal dynamics to describe rain streaks. Zhang et al.~\cite{zhang2006rain} exploit temporal and chromatic properties of rain in videos. Bossu et al.~\cite{bossu2011rain} use the histogram of orientations of rain streaks to detect them. And the details of other methods including~\cite{barnum2010analysis,bossu2011rain,santhaseelan2015utilizing} are addressed in~\cite{tripathi2014removal}. 

\subsection{Single Image De-raining}
Compared with video-based de-raining, single image de-raining is a challenging problem, since there is no temporal information in single images. Some single-image de-raining methods consider de-raining as a layer separation problem. Chen et al.~\cite{chen2013generalized} propose a generalized low rank model, and model rain streaks as a low rank layer. Kang et al.~\cite{kang2012automatic} separate high frequency parts of rainy images into background layer and rain layer.~\cite{Luo2015Removing} separates rain streaks from background by a discriminative sparse coding method. Kim et al.~\cite{Kim2014Single} detect and remove rain streaks via nonlocal mean filters. Moreover, a recent work proposed by ~\cite{Li2016Rain} separates rain streaks by exploiting Gaussian mixture models. 

Recently, several deep learning-based methods have shown promising results. Fu et al. \cite{fu2017clearing,fu2017removing} introduce deep learning to remove rain streaks. They combine a convolutional neural network with some traditional methods to separate images by frequency, and map high frequency parts to the rain streaks layer. Zhu et al.~\cite{zhu2017joint} propose a de-raining method by combining three different kinds of image priors. Yang et al.~\cite{yang2017deep} design a deep recurrent dilated network for de-raining. Zhang et al.~\cite{zhang2017image} use conditional generative adversarial network (conditional-GAN) to remove rain streaks and they utilize perceptual loss to ensure better quality of image reconstruction. Zhang et al.~\cite{zhang2018density} further use a pre-classifier to determine rain-density for de-raining. Li et al.~\cite{li2018recurrent} design a novel multi-stage CNN including several parallel sub-networks that are made aware of different scales of rain streaks for de-raining. 

However, owing to the interference of rain streaks, the reconstructed image will be too dim or saturated after de-raining. These methods usually pay more attention to remove rain streaks, leading to damaged color and textures. Although some methods address the problem and add a fidelity term to loss functions or use a classifier to distinguish different rain density in different conditions for better reconstruction, they still cause residual rain streaks in results. Adding fidelity to loss functions usually increases the difficulty of optimization and the classifier trained by a simple synthetic dataset is hard to handle the rain density in real complex conditions. 

\section{Dual-Task De-Raining Network (DTDN)}

\begin{figure*}[htp] 
\centering 
	\includegraphics*[width=38em]{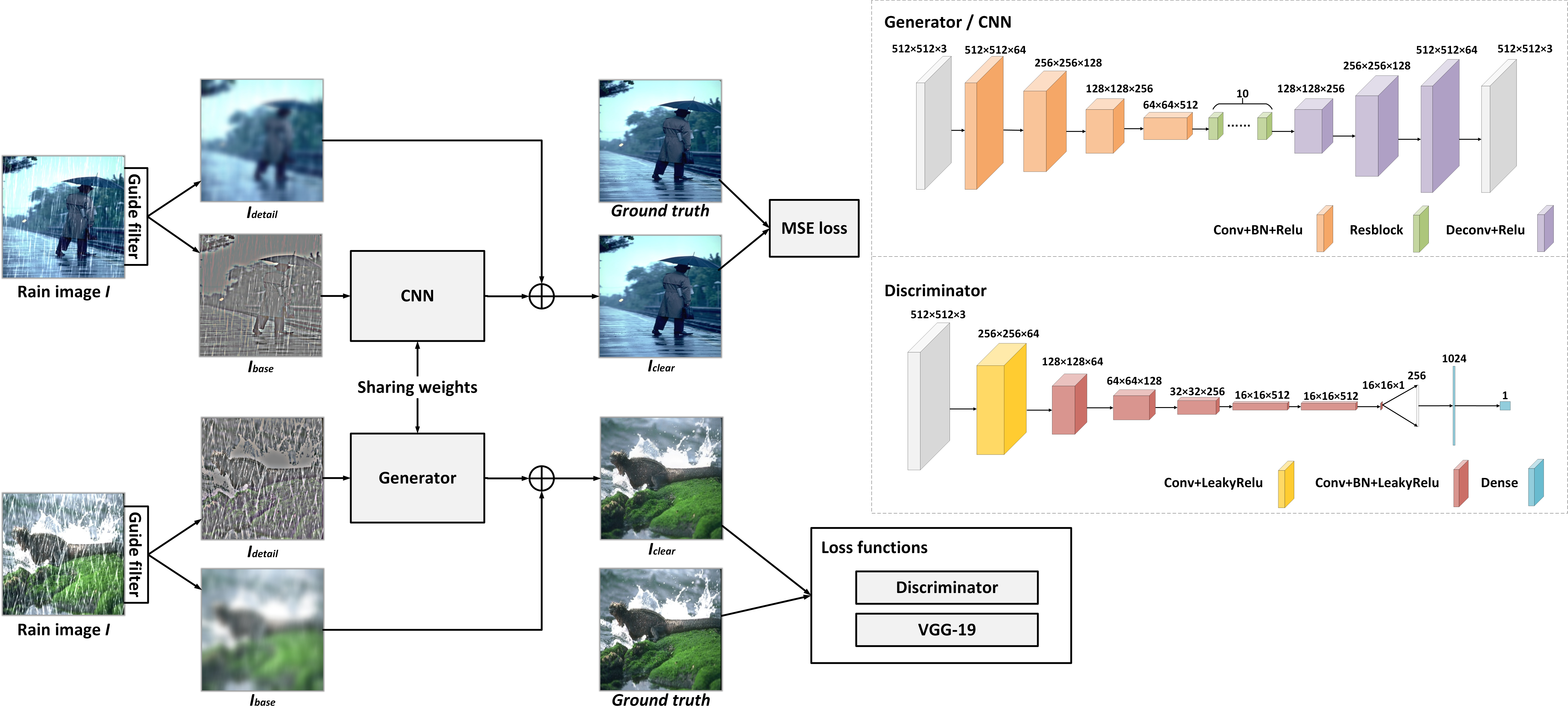}
\caption{The framework of DTDN. After training, given an input image with rain streaks, we only need to feed it to the generator of DTDN-GAN and an image without rain streaks can be generated through the following three steps. Firstly, a guide filter~\cite{he2013guided} is used to obtain $I_{base}$ and $I_{detail}$ from a rainy image $I$. Secondly, we compute $I^\prime_{detail}$=Generator($I_{detail}$) by sending $I_{detail}$ to generator. Finally, we add $I^\prime_{detail}$ and $I_{base}$ to obtain $I_{clear}$ where rain streaks are removed.} \label{DTDN}
\end{figure*}

We design a dual-task de-raining network (DTDN) that is composed of two sub-networks: a generative adversarial network~\cite{goodfellow2014generative} (DTDN-GAN) used for removing rain streaks, and a convolutional neural network (DTDN-CNN) for retaining realistic details of original images. DTDN-CNN shares the same weights and structure with the generator of DTDN-GAN, but they use different loss functions. They are trained individually and iteratively, based on different training data. The structure of DTDN is illustrated in Figure~\ref{DTDN}.

\subsection{DTDN-GAN}

DTDN-GAN is the first sub-network that uses a generator to remove rain streaks from an image and uses a discriminator to distinguish generated images from real images. Following the previous works~\cite{zhang2017image, zhang2018density}, the loss function of DTDN-GAN adopts the combination of content loss and adversarial loss. The content loss is designed via VGG~\cite{simonyan2014very} that is used to optimize the distance between ground truth and rainy image in feature space of VGG. Intuitively, the content loss and adversarial loss are fit for removing structural rain streaks, mainly because they are designed for classification and they are more sensitive to structural information of objects~\cite{zeiler2014visualizing}. 
The whole loss function of DTDN-GAN, denoted as perceptual loss, is designed as follows:
\begin{equation}\label{lam}
perceptual~loss = content\_loss + \lambda adversarial\_loss,
\end{equation}
where $\lambda$ is a hyper-parameter to control the balance between the two losses. The content loss is
\begin{equation}
\Vert VGG(I_{truth}) - VGG(I_{clear})\Vert_2,
\end{equation}
where the $VGG$ is a convolutional neural network (CNN), $VGG(\ast)$ is the feature map of $\ast$ by $VGG$ network, and $\Vert \ast \Vert_2$ is L-2 norm of matrix.
The adversarial loss is
\begin{equation}
\sum \left[log(D(I_{truth})) + log(1-D(I_{clear}))\right],
\end{equation}
where $D$ is discriminator, $G$ is generator and $I_{truth}$ is the ground truth for $I_{clear}$.

\subsection{DTDN-CNN}

DTDN-CNN is the second sub-network of DTDN for the reconstructed image from DTDN-GAN to preserve color and textures and remove artifacts generated by perceptual loss (the details are in~\cite{johnson2016perceptual}). Previously, the two problems are solved by subjoining a fidelity term (e.g., previous de-raining works ~\cite{zhang2017image, zhang2018density} use $loss~function=perceptual~loss+\lambda MSE$ where $\lambda$ is a hyper-parameter to control the balance between the two tasks). However, the method usually causes damaged color or residual rain streaks. The reasons why the results are unsatisafied will be analyzed in details in Section~\ref{sec:cwom}.

With intention to remove artifacts as well as preserve color and texture details, we add the second sub-network DTDN-CNN, which shares the same structure and weights with DTDN-GAN. DTDN-CNN adopts the mean square error (MSE) as its loss function that is more sensitive to pixel level information. The loss is designed as follows,
\begin{equation}
MSE = \Vert I_{truth} - I_{clear}\Vert_2.
\end{equation}

DTDN-GAN and DTDN-CNN are trained alternatively to solve the whole optimization problem.

\subsection{The De-raining Algorithm}

DTDN-GAN and DTDN-CNN use specific loss functions for serving two tasks of de-raining, respectively. We design a training method by adopting two measures to make DTDN-GAN focus on removing rain streaks, and make DTDN-CNN focus on recovering details from original images. Firstly, DTDN-GAN adopts a greater learning rate than DTDN-CNN, since we need to ensure the DTDN-GAN converges before DTDN-CNN. That means removing rain streaks is our main task. Secondly, the training data of DTDN-GAN only include the images with more complex rain streaks to make GAN have a stronger de-raining ability; DTDN-CNN uses all training images including both complex and simple rain streaks to make CNN obtain better effectiveness for recovering details of images. The algorithm is shown in Table~\ref{tab1}, where the `.heavy' denotes training data with heavy rain streaks, the `.all' denotes using all training data, $I_{truth}$ represents ground truth for synthetic rainy images and other symbols are explained in Figure~\ref{DTDN}.

\begin{figure}[htp] 
\centering 
	\subfigure[]{ \label{a} 
	\includegraphics*[width=0.23\linewidth]{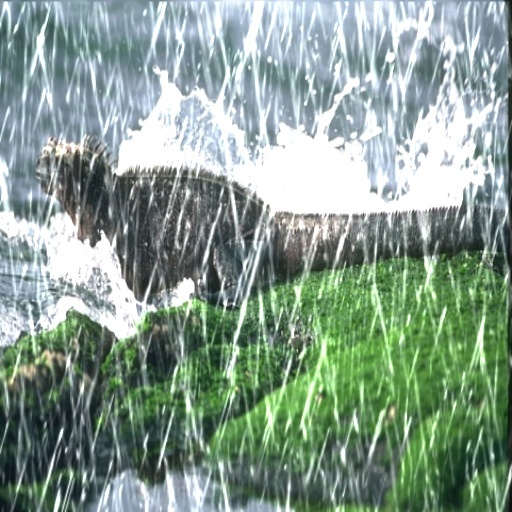}} 
	\subfigure[]{ \label{b} 
	\includegraphics*[width=0.23\linewidth]{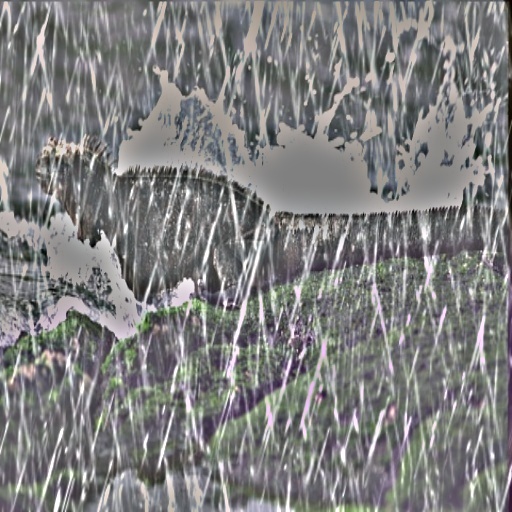}} 
	\subfigure[]{ \label{c} 
	\includegraphics*[width=0.23\linewidth]{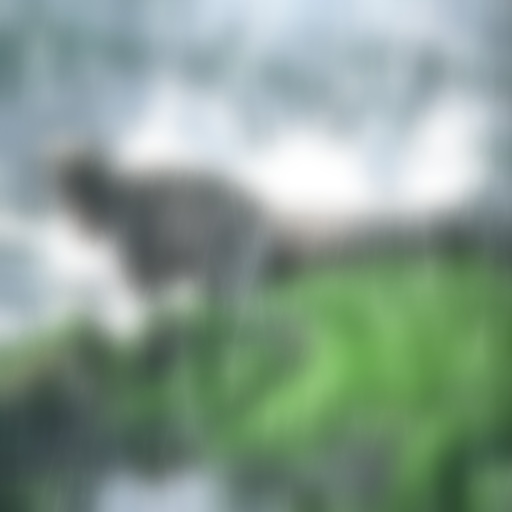}} 
	\subfigure[]{ \label{d} 
	\includegraphics*[width=0.23\linewidth]{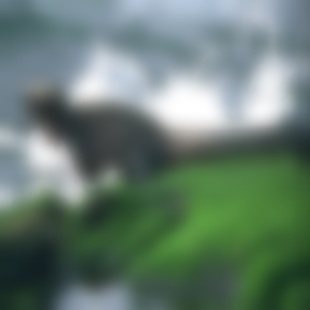}} 
\caption{(a) a rainy image; (b) $I_{detail}$ of the rainy image; (c) $I_{base}$ of the rainy image; (d) $I_{base}$ filtered from~$I_{truth}$.} \label{fig3}
\end{figure}

\begin{table}[h]
\centering
\caption{The training algorithm of DTDN.}\label{tab1}
\renewcommand\arraystretch{0.8}
\begin{tabular}{l}
\toprule[2pt]
DTDN.GAN.Generator.learning rate = 0.001 \\
DTDN.CNN.learning rate = 0.0001 \\
\midrule[1pt] 
cycle begin:\\
$\qquad$ $I$ = read randomly a batch from Dataset.heavy()\\
$\qquad$ $I_{base}$ = Guide filter($I$)\\
$\qquad$ $I_{detail} = I - I_{base}$\\
$\qquad$ $I_{detail}^\prime$ = DTDN.GAN.Generator($I_{detail}$)\\
$\qquad$ $I_{clear}$ = $I_{detail}^\prime$ + $I_{base}$ \\
$\qquad$ train DTDN.GAN.discriminator($I_{truth}$, $I_{clear}$)\\
$\qquad$ train DTDN.GAN.generator($I_{truth}$, $I$)\\
cycle end\\
$I$ = randomly read a batch from Dataset.all()\\
$I_{base}$ = Guide filter($I$)\\
$I_{detail} = I - I_{base}$\\
$I_{detail}^\prime$ = DTDN.GAN.Generator($I_{detail}$)\\
$I_{clear}$ = $I_{detail}^\prime$ + $I_{base}$\\
train DTDN.CNN($I_{truth}$,$I$)\\
DTDN.CNN Shares weights with DTDN.GAN \\
\bottomrule[2pt]
\end{tabular}
\end{table}

The training algorithm has two steps. The first step is to train DTDN-GAN. At the beginning of training, a batch of heavy rainy images from a dataset are taken randomly, a guide filter is used to obtain the filtered image $I_{base}$, and $I_{detail}$(=$I$-$I_{base}$) is further computed. Using $I_{detail}$ and $I_{base}$ as input, we further train the discriminator and generator of DTDN-GAN. The reason why we prefer to use $I_{detail}$ rather than the original rainy image $I$ is that using $I_{detail}$ as input can simplify the process of image reconstruction. As shown in Figure~\ref{fig3}, $I_{detail}$ contains the majority of rain streaks of $I$ whose $I_{base}$ is similar to that of $I_{truth}$.

Furthermore, we obtain fake $I$ (the name for standard GAN), named as $I_{clear}$ by using the forward propagation of generator. According to the loss between $I_{clear}$ and the label $I_{truth}$, we optimize discriminator and generator of DTDN-GAN by using the same way as standard GAN.

After training DTDN-GAN, the second step is to train DTDN-CNN. DTDN-CNN uses a smaller learning rate, more comprehensive training data and a finer-grained loss function to recover image details.

\subsection{Data Enrichment}

\begin{figure}[htp] 
\centering 
	\subfigure[An object in raining]{ \label{intro11} 
	\includegraphics*[width=0.45\linewidth]{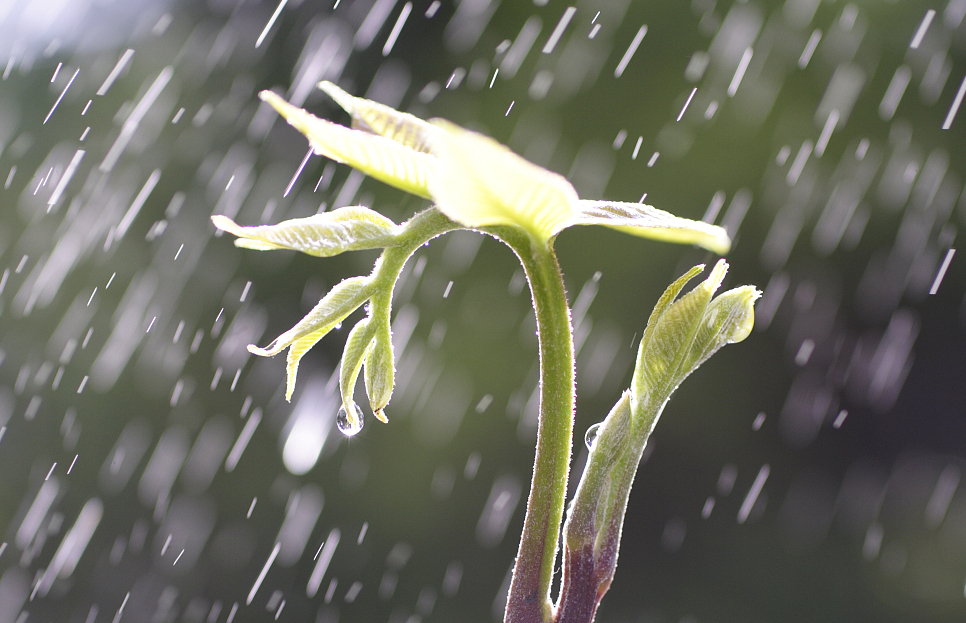}} 
	\subfigure[A common picture]{ \label{intro12} 
	\includegraphics*[width=0.45\linewidth]{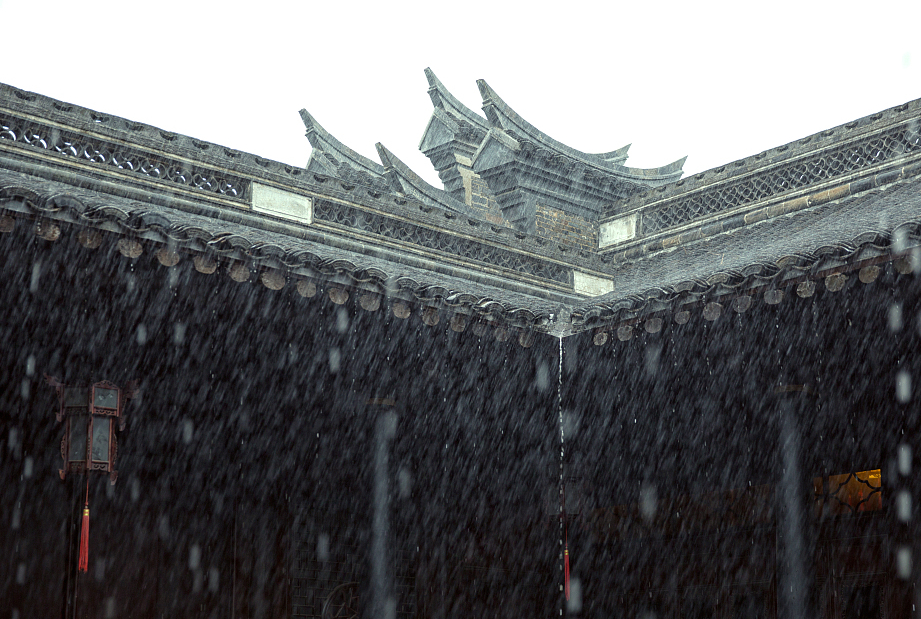}} 
\caption{(a) has rain streaks in different shapes, strengths and directions, mainly caused by the effect of device. (b) has different rain streaks, due to wind.} \label{intro1}
\end{figure}

Although there exist several datasets of rainy images, they only contain single-type rain streaks in each image. In Figure~\ref{intro1}, different types of rain streaks often appear in the same image in the real world. The phenomenon may be caused by physical property of imaging devices or some natural factors like wind or terrain, which are inevitable. Concretely, in Figure~\ref{intro11}, due to depth of field (DOF) to cameras, there are two different types of rain streaks emerging at background and foreground. And in Figure~\ref{intro12}, owing to the wind and the distance to the camera, rain streaks also show different forms. Such complex rain streaks are intractable for trained models based on previous training sets. Hence, we propose a technique to enrich existing datasets.

Facing more complex conditions like Figure~\ref{intro1}, we should enrich training data by adding more types of rain streaks with more directions into rainy images. Since $RainTrainL$~\cite{yang2017deep} contains some images of pure rain streaks which do not have any background, we add these pure rain streaks into the images in the datasets to be enriched. Specifically, we divide rain streaks into three categories depending on their directions and numbers, select 0$\sim$3 types of rain streaks randomly from $RainTrainL$ and add them to the dataset. Compared with those existing datasets of rainy images, each image of the dataset after enrichment contains more types of rain streaks with more directions, which makes de-raining models more capable of handling complex rainy conditions.

\subsection{Discussion} \label{sec:cwom}
This section will address theoretically the significance of the proposed training algorithm, which trains two sub-networks alternatively based on two different data sources, respectively.

\textbf{1. Why do we combine the perceptual loss function and MSE loss function ?} If we merely use MSE loss function, the results will be usually too smooth, and if we only use perceptual loss function, some artifacts will emerge in the results. Therefore, we combine the two loss functions to reach a compromise. Mathematically, the compromise requires the convergences of the two loss functions simultaneously.

\textbf{2. Why do we train the two sub-networks of DTDN alternatively rather than train them together?} Since we employ a data enrichment technique to complicate training data, it makes training become harder. Finding an efficient method is necessary to reduce the optimization cost. We begin with analyzing the problem of subjoining MSE loss function to perceptual loss as a whole. Here is its loss function:
\begin{equation}\label{loss}
l=\parallel VGG\lbrack f(I;\theta) \rbrack -VGG(I_{truth})\parallel_2+ \lambda\parallel  f(I;\theta)  -I_{truth}\parallel_2
\end{equation}
where $f(I;\theta)=I_{clear}$ and $f$ represents a neural network. When the network encounters a local minimum, the following equation holds (For convenience, we directly use squared terms for $L_2$ norm in Equation~(\ref{loss})).
\begin{equation}\label{pequ}
\begin{aligned}
\frac {\partial f(I;\theta)} {\partial \theta} \lbrack \frac {\partial VGG(I_{clear})} {\partial I_{clear}}\lbrack VGG(I_{clear})-VGG(I_{truth})\rbrack~&+ \\ \lambda (I_{clear}-I_{truth})\rbrack& =0
\end{aligned}
\end{equation}
Subjoining MSE to perceptual loss directly makes the optimization objective more complex and leads to many unreasonable local optima. A possible local minimum that seems to be unrelated to the global minimum is as following:
\begin{equation}\label{errorsolu}
\left\{
\begin{aligned}
&\mid \frac {\partial VGG(I_{clear})} {\partial I_{clear}} \mid VGG(I_{clear})-\lambda I_{clear} =0\\ &~\lambda I_{truth}- \mid \frac {\partial VGG(I_{clear})} {\partial I_{clear}} \mid VGG(I_{truth})=0
\end{aligned}
\right.
\end{equation}
when the signs of $(VGG(I_{clear})-VGG(I_{truth}))$ and $(I_{clear}-I_{truth})$ are the same, the sign of $\frac {\partial VGG(I_{clear})} {\partial I_{clear}}$ is different from theirs. When Equations~(\ref{errorsolu}) holds, the Equation~(\ref{pequ}) also holds, but the two loss functions do not converge to their minima, diverging from our optimization objective. That means, using the previous training way may lead to interference between the two loss functions. It increases the uncertainty of convergence to expected solutions.

Whereas, when our training converges, the following two equations must be satisfied at the same time, i.e., only if the two sub-networks converge, will the entire network converge.
\begin{equation}\label{coequ}
\left\{
\begin{aligned}
&\frac {\partial f(I;\theta)} {\partial \theta}   \frac {\partial VGG(I_{clear})} {\partial I_{clear}} \lbrack VGG(I_{clear})- VGG(I_{truth})\rbrack=0
\\& \frac {\partial f(I;\theta)} {\partial \theta}(I_{clear}-I_{truth}))=0 
\end{aligned}
\right.
\end{equation}
The solutions to Equations~(\ref{coequ}) are

\begin{equation}\label{solu1}
\frac {\partial f(I;\theta)} {\partial \theta}=0 \quad  \text{(local minimum)}
\end{equation}
or
\begin{equation}\label{solu2}
I_{clear}=I_{truth}\quad\text{(global minimum)}
\end{equation}
When the Solution~(\ref{solu1}) holds, Equations~(\ref{coequ}) also hold. When Solution~(\ref{solu2}) holds, we can deduce $VGG(I_{clear})-VGG(I_{truth})=0$, then Equations~(\ref{coequ}) hold naturally. In other words, the cost of solving Equations~(\ref{coequ}) is equal to the cost of training a network with an MSE function. We spend the optimization cost of training with one loss function, obtaining the effectiveness of training with two loss functions.

\textbf{3. why do we train the two sub-networks on different training sets ?} Based on Equations~(\ref{coequ}), to make its optimizer reach the global minimum, we exert a momentum to make the optimization leave local minima ($\frac {\partial f(I;\theta)} {\partial \theta}=0$). The momentum is produced by training the two sub-networks on two different data sources which are from different distributions. In details, the solutions corresponding to Equations~(\ref{coequ}) become:

\begin{equation}\label{newg}
\begin{aligned}
&\left\{
\begin{split}
&\frac {\partial f(I;\theta)} {\partial \theta}=0 \\ &\frac {\partial f(I^h;\theta)} {\partial \theta}=0
\end{split}
\right.~(\ref{newg}\text{a})\quad
&\left\{
\begin{split}
&\frac {\partial f(I;\theta)} {\partial \theta}=0 \\ &I_{clear}^h=I_{truth}^h
\end{split}
\right.~~(\ref{newg}\text{b})&\\ 
&\left\{
\begin{split}
&I_{clear}=I_{truth} \\&\frac {\partial f(I^h;\theta)} {\partial \theta}=0
\end{split}
\right.(\ref{newg}\text{c})\quad
&\left\{
\begin{split}
&I_{clear}=I_{truth} \\ &I_{clear}^h=I_{truth}^h
\end{split}
\right.~(\ref{newg}\text{d})&
\end{aligned}
\end{equation}
where $I$, $I^h$ represent images from dataset.all and dataset.heavy, respectively. The Solution~(\ref{newg}) includes four types of solutions that are obtained by bringing $I$, $I^h$ into Equations~(\ref{solu1}) and~(\ref{solu2}), repectively. 

Compared with the Solution~(\ref{solu1}), Solution~(\ref{newg}a) has a lower possibility of occurrence, because $p(\frac {\partial f(I;\theta)} {\partial \theta}=0 \text{ and } \frac {\partial f(I^h;\theta)} {\partial \theta}=0) < p(\frac {\partial f(I;\theta)} {\partial \theta}=0)$ where $I$ and $I^h$ are from two different distributions, i.e., $p(\frac {\partial f(I^h;\theta)} {\partial \theta}=0|\frac {\partial f(I;\theta)} {\partial \theta}=0) < 1$. That means, when DTDN-GAN encounters a local optimum, the DTDN-CNN may help DTDN-GAN jump it out. In other words, the momentum to leave local minima is that DTDN-CNN and DTDN-GAN provide regularization constraints for each other. 

\textbf{4. How to obtain the solution as expected ?} Solution~(\ref{newg}b) is the result as expected, which uses perceptual loss function to remove rain streaks and uses MSE loss function to refine color and textures. That means, DTDN-GAN converges to the global minimum and DTDN-CNN converges to a local minimum. Although Solution~(\ref{newg}d) seems to be more acceptable, it represents the effect of training a network with only MSE, since DTDN-CNN uses larger training set than DTDN-GAN.

Our training algorithm increases the possibility of Solution~(\ref{newg}b) of occurrence by reducing those of the other three solutions. Firstly, the algorithm reduces the probability of Solution~(\ref{newg}a) happening by training the two sub-networks on different training sets (details in the third question). Secondly, since we set a lower learning rate for DTDN-CNN, DTDN-GAN should converge before DTDN-CNN converges generally, which reduces possibility of Solution~(\ref{newg}c). Finally, convergences to global minima of the two sub-networks (Solution~(\ref{newg}d)) simultaneously is a small probability event. 

This training algorithm is inspired by the branch and bound technique. The feasible solution space is segmented into smaller subsets and the expected solutions can be obtained by reducing the occurrence of other solutions ((\ref{newg}a), (\ref{newg}c) and (\ref{newg}d)).

\textbf{Summary.} To achieve the solution as expected, our training algorithm adopts three measures. Firstly, the algorithm trains two sub-networks of DTDN alternatively rather than subjoining them, aiming to reduce uncertainty of solutions, (the details in the second questions). Secondly, it trains two sub-networks on different training sets, segmenting solutions into four types, which makes the bound of solution space more clear (the details in the third question). Finally, it uses a lower learning rate for DTDN-CNN and trains the two subnetworks alternatively, reducing the occurrence probability of the other three solutions (\ref{newg}a), (\ref{newg}c) and (\ref{newg}d), such that the expected solution~(\ref{newg}b) can be obtained with a greater probability (the details in the fourth question).


\section{Experiments}
We compare DTDN with the recent state-of-the-art methods: CNN~\cite{fu2017clearing}, DNN~\cite{fu2017removing}, JORDER~\cite{yang2017deep}, ID-CGAN~\cite{zhang2017image}, DID-MDN~\cite{zhang2018density} and RESCAN~\cite{li2018recurrent}, using Peak Signal to Noise Ratio (PSNR), Structural Similarity Index (SSIM)~\cite{wang2004image} and Universal Quality Index (UQI)~\cite{wang2002universal}. 

\subsection{Datasets}
In general, an entire dataset consists of training set and testing set. We use the enrichment technique to enhance the two sets of~\cite{zhang2018density} denoted as $CRain$ and $CRain\textrm{-}test$. Further, we append a part of testing set of~\cite{fu2017removing} to $CRain\textrm{-}test$. 

\textbf{Testing set.} We use $CRain\textrm{-}test$, $Test1$ and $Test2$~\cite{zhang2018density} to evaluate different methods. Moreover, to evaluate the generalization capacity of different models in the real world, we create a Real-World $(RW)$ test set containing 30 simple or complex rainy images.

\textbf{Training set.} $CRain$ is used to train DTDN, RESCAN and JORDER. The other methods including DID-MDN, CNN, DDN and ID-CGAN directly use their own weight files.

\subsection{Training Details}

Before training, we resize all images into 512$\times$512. Our model uses Adam~\cite{Kingma2014Adam} to optimize its weights where $\beta_1$ is 0.9, $\beta_2$ is 0.99 and the learning rates of DTDN-GAN and DTDN-CNN are 0.001 and 0.0001, respectively. The local window radius and regularization parameters of guide filter are both 10. The batch size of training data is 4 and the epoch is 60000. $\lambda$ in Eq.(\ref{lam}) is 100. More details are shown in our code.

\subsection{Ablation Study}

We do several ablation experiments to evaluate the effect of the proposed method.

\begin{figure}[!h] 
\centering 

\subfigure{ \label{rain_1}
\begin{minipage}[c]{\linewidth}\centering
\parbox[t]{0.28\linewidth}{\includegraphics*[width=1\linewidth]{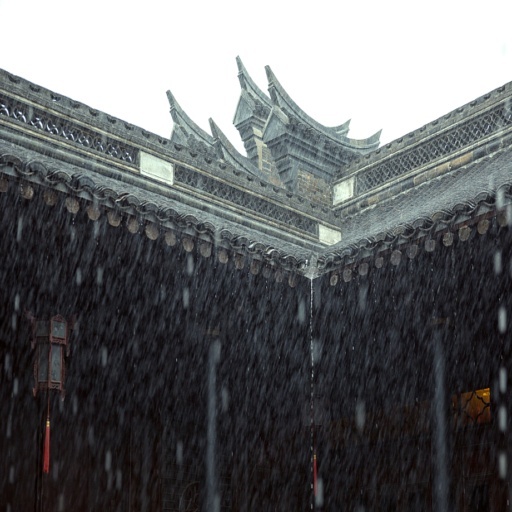}} 
\parbox[t]{0.28\linewidth}{\includegraphics*[width=1\linewidth]{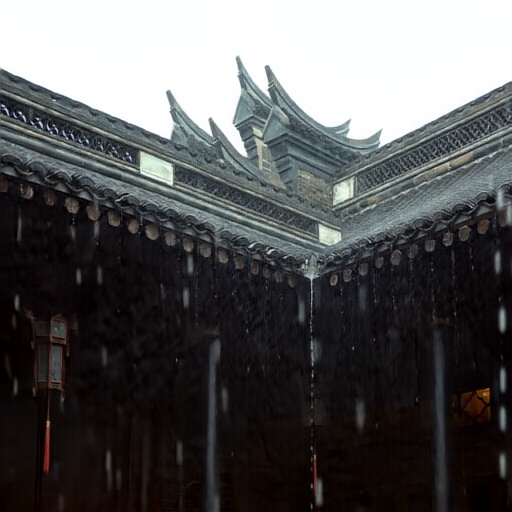}} 	
\parbox[t]{0.25\linewidth}{\includegraphics*[width=1\linewidth]{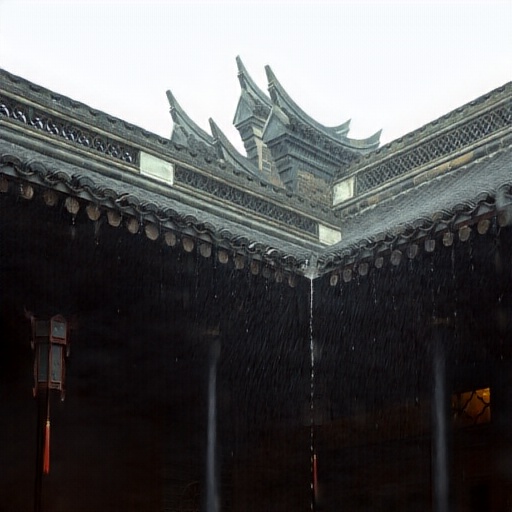}} 		
\end{minipage}}	

\subfigure{ \label{rain_2}
\begin{minipage}[c]{\linewidth}\centering
\parbox[t]{0.28\linewidth}{\includegraphics*[width=1\linewidth]{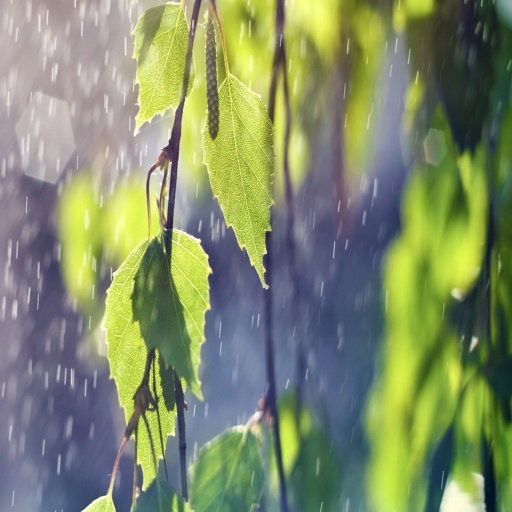}} 
\parbox[t]{0.28\linewidth}{\includegraphics*[width=1\linewidth]{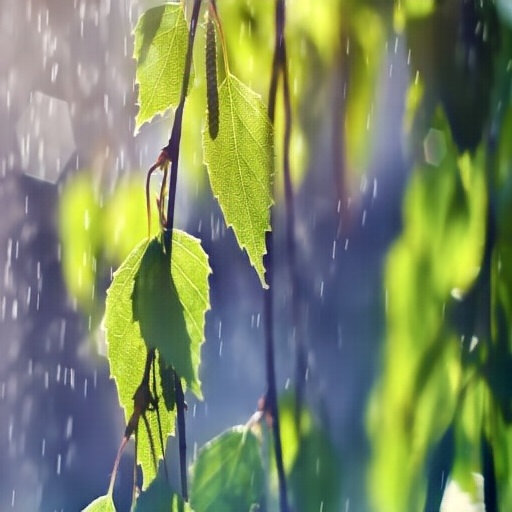}} 	
\parbox[t]{0.28\linewidth}{\includegraphics*[width=1\linewidth]{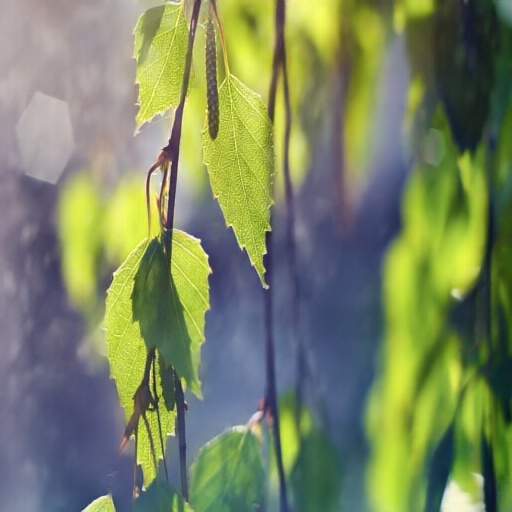}} 	
\end{minipage}}

\parbox[center]{0.28\linewidth}{\centering(a) }
\parbox[center]{0.28\linewidth}{\centering(b) }
\parbox[center]{0.28\linewidth}{\centering(c) }

\caption{Comparison of DTDN with the training set of $CRain$ and DTDN with the training set of \cite{zhang2018density} on real rainy test images. (a) real rainy images, (b) the results of DTDN trained on the dataset of \cite{zhang2018density}, (c) the results of DTDN on $CRain$.} \label{datasetvi}
\end{figure}

Firstly, we explore whether or not the data enrichment is effective for training de-raining algorithm. We train our model (DTDN) based on the training data of~\cite{zhang2018density} and the dataset after enrichment (denoted as $CRain$). The comparison results on $Test1$ are shown in Table~\ref{tab5} and the results on $RW$ are shown in Figure~\ref{datasetvi}. The DTDN that uses $CRain$ obtains comparatively lower metric values because $CRain$ contains several types of rain streaks with different directions, which may influence the reconstruction metric values. Nonetheless, the visual performance of DTDN using $CRain$ outperforms that trained on the dataset before enrichment. Though the way of data enrichment decreases the metrics slightly, its visual effect is better. In other words, owing to data enrichment, our model is capable of dealing with more complex rainy conditions.
\begin{table}[!h]
\newcommand{\tabincell}[2]{\begin{tabular}{@{}#1@{}}#2\end{tabular}}
\centering
\caption{The results of DTDN using training set of $CRain$ and DTDN with the training set of \cite{zhang2018density} on $Test1$.}\label{tab5}
\begin{tabular}{|c|c|c|c|}
\hline
 &  \tabincell{c}{DTDN using the \\ previous training set} & \tabincell{c}{DTDN using the training\\ set of $CRain$ }\\
\hline
PSNR &  \textbf{29.21}	&  29.01  \\
\hline
SSIM &  \textbf{0.9235}	&  0.9234 \\
\hline
UQI &   \textbf{0.9434}	&  0.9432  \\
\hline
\end{tabular}
\end{table}

Secondly, we evaluate the effectiveness of different modules in our method by conducting the following experiments:

\begin{itemize}
\item[$\bullet$] Experiment 
\uppercase\expandafter{\romannumeral1}: Removing DTDN-CNN from DTDN and only training DTDN-GAN, i.e., only using perceptual loss for the entire network.
\item[$\bullet$] Experiment 
\uppercase\expandafter{\romannumeral2}: Removing DTDN-CNN and only training DTDN-GAN with $perceptual~loss + MSE$ as its loss function.
\item[$\bullet$] Experiment 
\uppercase\expandafter{\romannumeral3} : Training standard DTDN.
\end{itemize}

\begin{figure}[!h] 
\centering 

\subfigure{ \label{fruit}
\begin{minipage}[c]{\linewidth}\centering
\parbox[t]{0.23\linewidth}{\includegraphics*[width=1\linewidth]{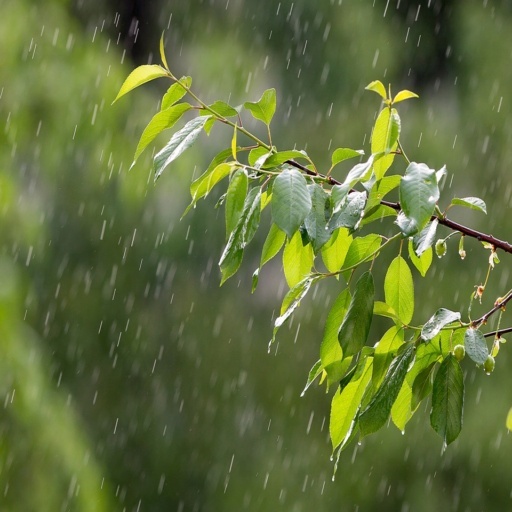}} 
\parbox[t]{0.23\linewidth}{
\includegraphics*[width=1\linewidth]{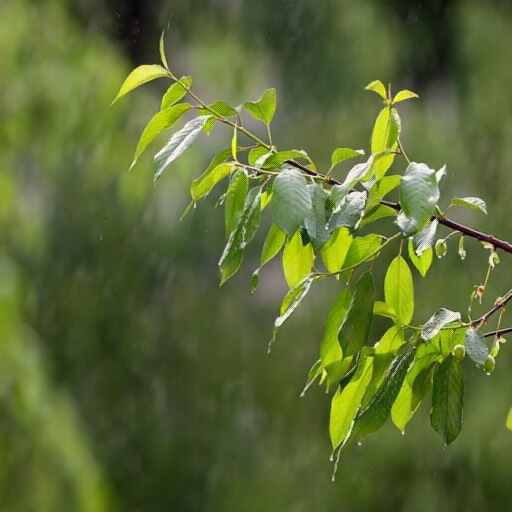}} 	
\parbox[t]{0.23\linewidth}{
\includegraphics*[width=1\linewidth]{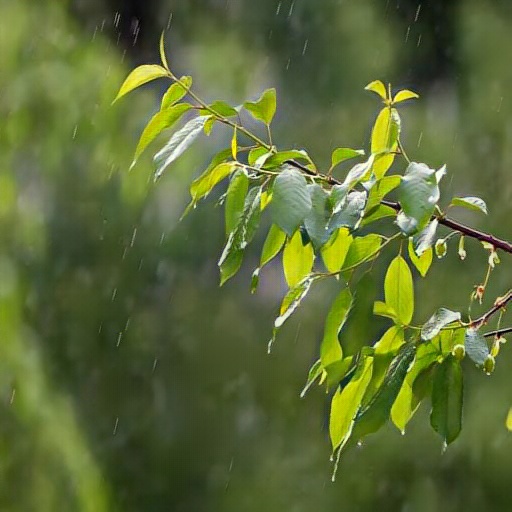}} 	
\parbox[t]{0.23\linewidth}{
\includegraphics*[width=1\linewidth]{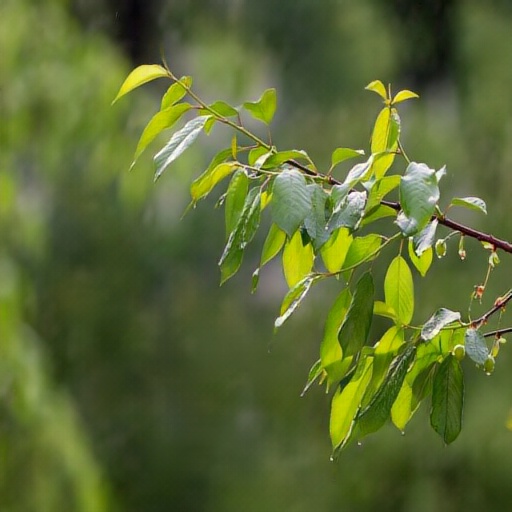}} 	
\end{minipage}}	

\subfigure{ \label{fruit}
\begin{minipage}[c]{\linewidth}\centering
\parbox[t]{0.23\linewidth}{\includegraphics*[width=1\linewidth]{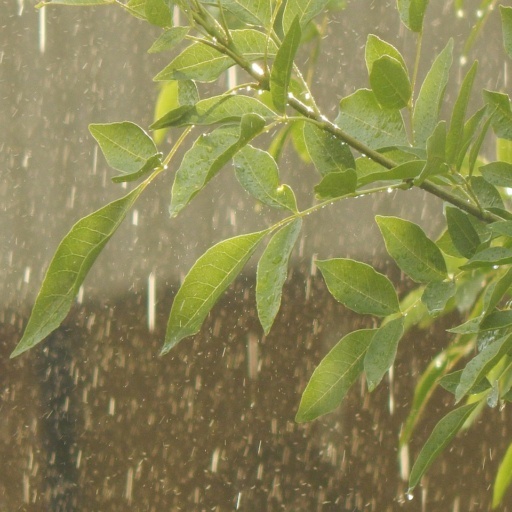}} 
\parbox[t]{0.23\linewidth}{
\includegraphics*[width=1\linewidth]{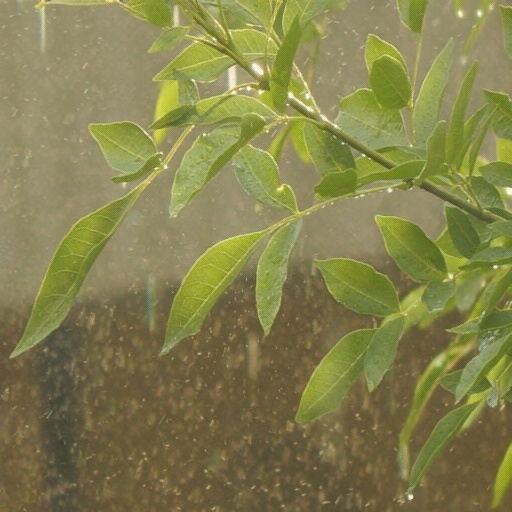}} 	
\parbox[t]{0.23\linewidth}{
\includegraphics*[width=1\linewidth]{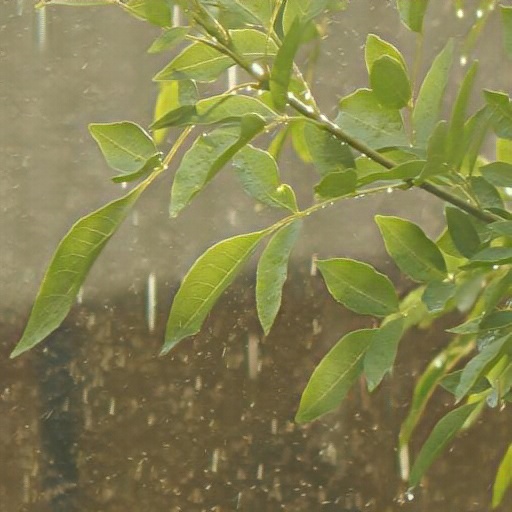}} 	
\parbox[t]{0.23\linewidth}{
\includegraphics*[width=1\linewidth]{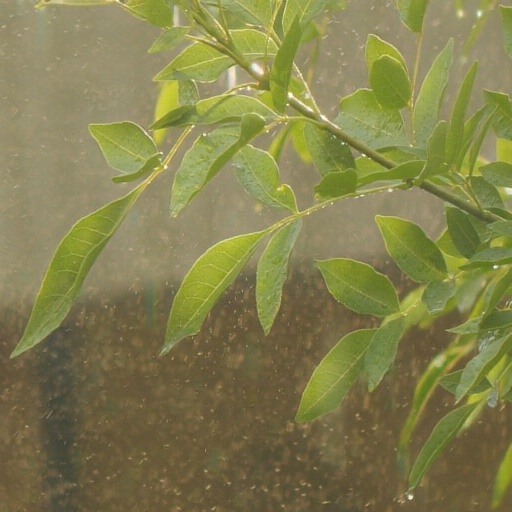}} 	
\end{minipage}}	

\subfigure{ \label{fruit}
\begin{minipage}[c]{\linewidth}\centering
\parbox[t]{0.23\linewidth}{\includegraphics*[width=1\linewidth]{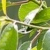}} 
\parbox[t]{0.23\linewidth}{
\includegraphics*[width=1\linewidth]{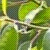}} 	
\parbox[t]{0.23\linewidth}{
\includegraphics*[width=1\linewidth]{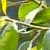}} 	
\parbox[t]{0.23\linewidth}{
\includegraphics*[width=1\linewidth]{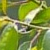}} 	
\end{minipage}}	

\parbox[center]{0.23\linewidth}{\centering(a) rainy images}
\parbox[center]{0.23\linewidth}{\centering(b) Experiment 
\uppercase\expandafter{\romannumeral1}}
\parbox[center]{0.23\linewidth}{\centering(c) Experiment 
\uppercase\expandafter{\romannumeral2}}
\parbox[center]{0.23\linewidth}{\centering(d) Experiment 
\uppercase\expandafter{\romannumeral3}}

\caption{Comparison of three experiments. The first and second row show the results of three experiments. The third row are the local amplifications of the results in the first row, where some artifacts from Experiment 
\uppercase\expandafter{\romannumeral1} and the results of Experiment 
\uppercase\expandafter{\romannumeral2} are over-smooth.} \label{mseloss}
\end{figure} 

From Experiments 
\uppercase\expandafter{\romannumeral1} and 
\uppercase\expandafter{\romannumeral2}, we can investigate the effect of using MSE. From Experiments 
\uppercase\expandafter{\romannumeral1} and 
\uppercase\expandafter{\romannumeral3}, we can evaluate the effect of DTDN-CNN. The difference between Experiments 
\uppercase\expandafter{\romannumeral2} and 
\uppercase\expandafter{\romannumeral3} is that Experiment \uppercase\expandafter{\romannumeral2} uses fixed ratio for two objective functions but Experiment \uppercase\expandafter{\romannumeral3} uses DTDN-CNN to coordinate them. From Experiments 
\uppercase\expandafter{\romannumeral2} and 
\uppercase\expandafter{\romannumeral3}, we can infer which network structure is more effective.

The average PSNR, SSIM and UQI values evaluated on $CRain\textrm{-}test$ are tabulated in Table~\ref{tabx} and the samples from $RW$ are shown in Figure~\ref{mseloss}. From Experiments 
\uppercase\expandafter{\romannumeral1} and 
\uppercase\expandafter{\romannumeral2}, when we add $MSE$ to loss functions of DTDN-GAN, the values of three metrics increase. Based on their local amplification, the increases of three metrics are mainly because adding $MSE$ to its loss function alleviates the artifacts generated from $perceptual~loss$. However, the results of Experiment \uppercase\expandafter{\romannumeral2} have residual artifacts and rain streaks. The visual effect is even worse than before adding the term. We infer that the two loss functions interfere with each other leading to unreasonable local minima. From Experiments 
\uppercase\expandafter{\romannumeral1} and 
\uppercase\expandafter{\romannumeral3}, the results show that DTDN-CNN not only increases the quality of image reconstruction but also improves the performance of de-raining. By local amplification, we can observe that the artifacts produced by perceptual loss disappear so that the evaluation metric values increase, which proves that DTDN-CNN can improve the performance of removing rain streaks on a fine-grained level. From Experiments 
\uppercase\expandafter{\romannumeral2} and 
\uppercase\expandafter{\romannumeral3}, the learnability of DTDN-CNN can coordinate the relationship between two mutually exclusive objectives and can improve the effectiveness of de-raining on both quality and quantity.

\begin{table}[!h]
\newcommand{\tabincell}[2]{\begin{tabular}{@{}#1@{}}#2\end{tabular}}                                                                                                                                                                  
\centering
\caption{The results of DTDN without MSE and standard DTDN on $CRain\textrm{-}test$.}\label{tabx}
\begin{tabular}{|c|c|c|c|}
\hline
 &  \tabincell{c}{Experiment 
\uppercase\expandafter{\romannumeral1}} & \tabincell{c}{Experiment 
\uppercase\expandafter{\romannumeral2}} & \tabincell{c}{Standard  DTDN} \\
\hline
PSNR &  27.24	& 27.36  & \textbf{28.37} \\
\hline
SSIM &  0.8797	& 0.9025 & \textbf{0.9183} \\
\hline
UQI &   0.9479	& 0.9491 & \textbf{0.9538} \\
\hline
\end{tabular}
\end{table}

\subsection{Results on Three Synthetic Datasets}
\begin{figure*}[!h] 
\centering 

\subfigure{ \label{small}
\begin{minipage}[c]{\linewidth}\centering
\parbox[t]{0.15\linewidth}{
\includegraphics*[width=1\linewidth]{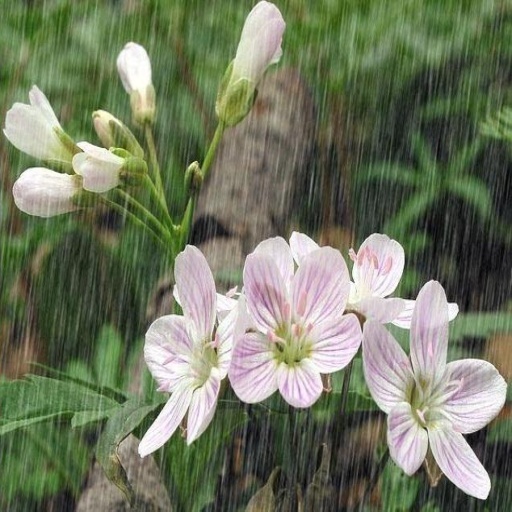}} 
\parbox[t]{0.15\linewidth}{
\includegraphics*[width=1\linewidth]{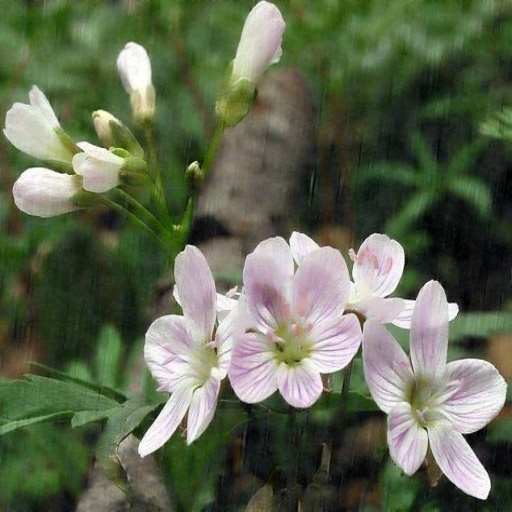}}  	
\parbox[t]{0.15\linewidth}{
\includegraphics*[width=1\linewidth]{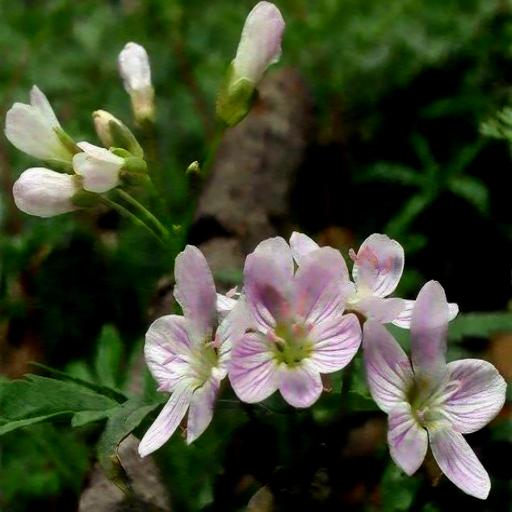}} 	
\parbox[t]{0.15\linewidth}{
\includegraphics*[width=1\linewidth]{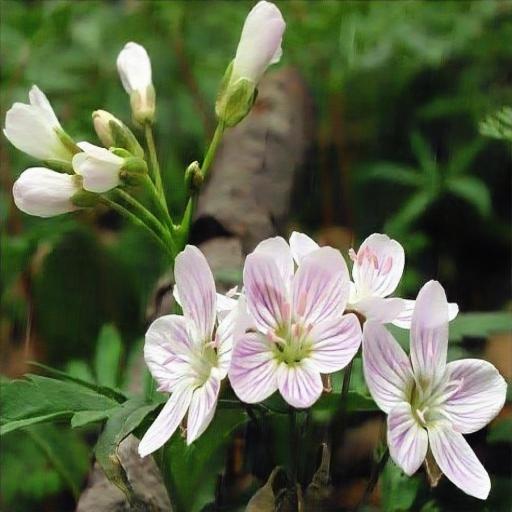}} 
\parbox[t]{0.15\linewidth}{
\includegraphics*[width=1\linewidth]{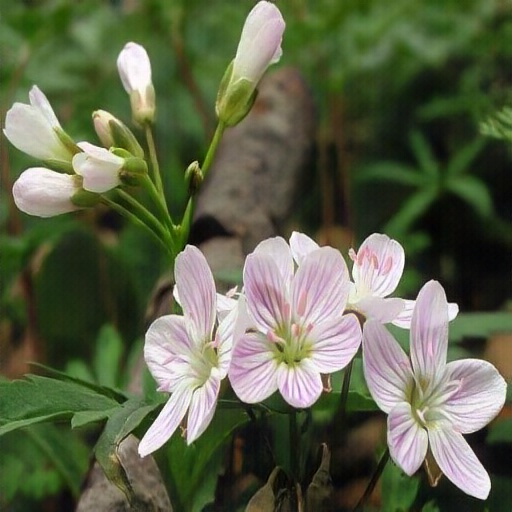}} 
\parbox[t]{0.15\linewidth}{
\includegraphics*[width=1\linewidth]{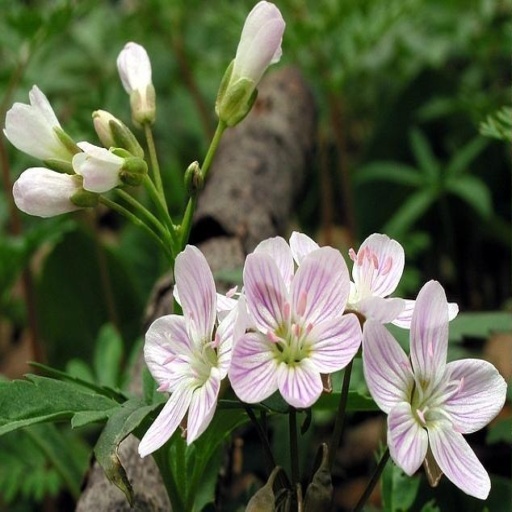}} 
\end{minipage}}	

\subfigure{ \label{6b}
\begin{minipage}[c]{\linewidth}\centering
\parbox[t]{0.15\linewidth}{
\includegraphics*[width=1\linewidth]{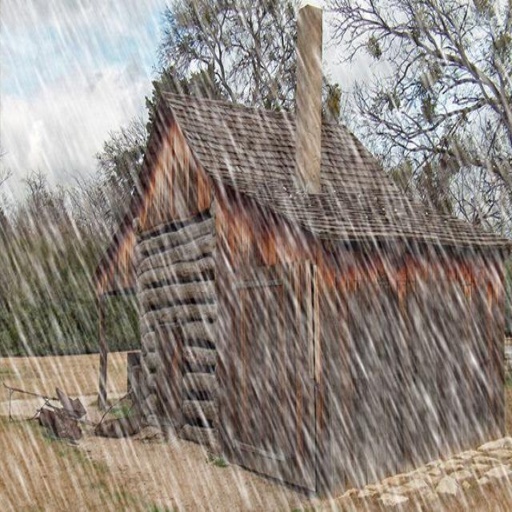}} 
\parbox[t]{0.15\linewidth}{
\includegraphics*[width=1\linewidth]{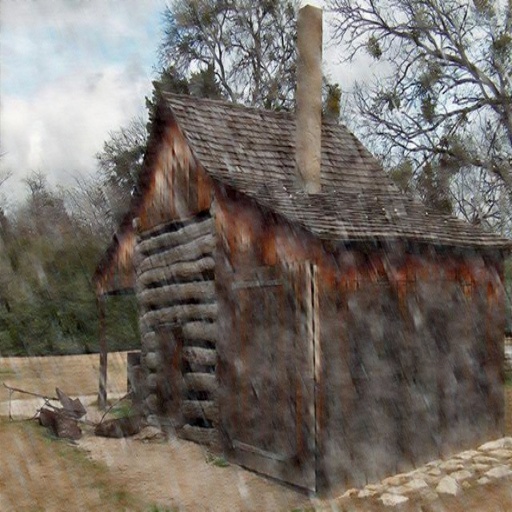}}  	
\parbox[t]{0.15\linewidth}{
\includegraphics*[width=1\linewidth]{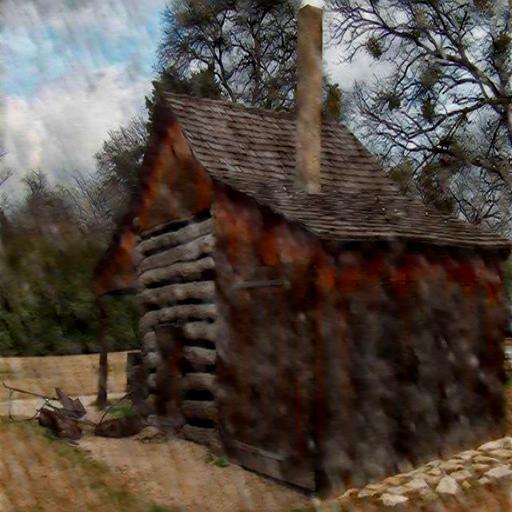}} 	
\parbox[t]{0.15\linewidth}{
\includegraphics*[width=1\linewidth]{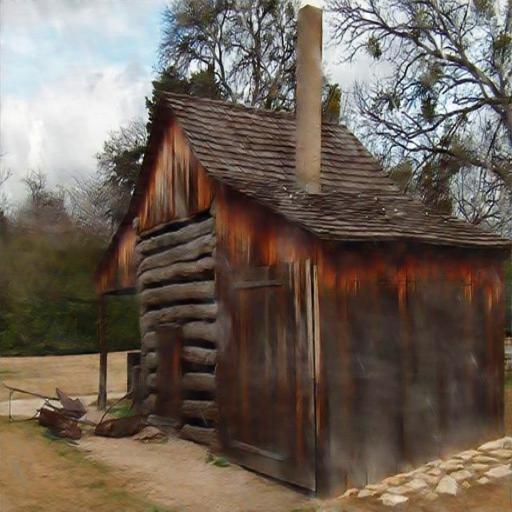}} 
\parbox[t]{0.15\linewidth}{
\includegraphics*[width=1\linewidth]{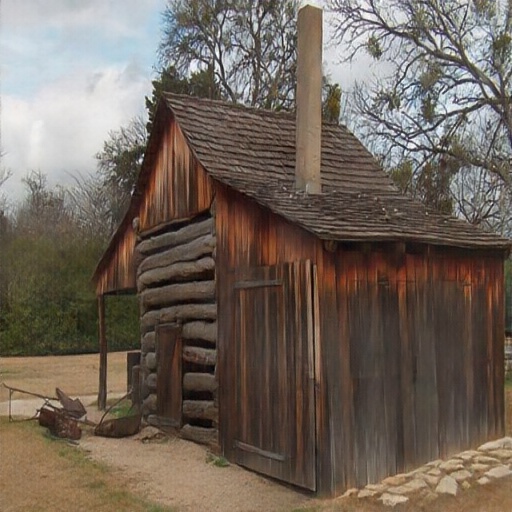}} 
\parbox[t]{0.15\linewidth}{
\includegraphics*[width=1\linewidth]{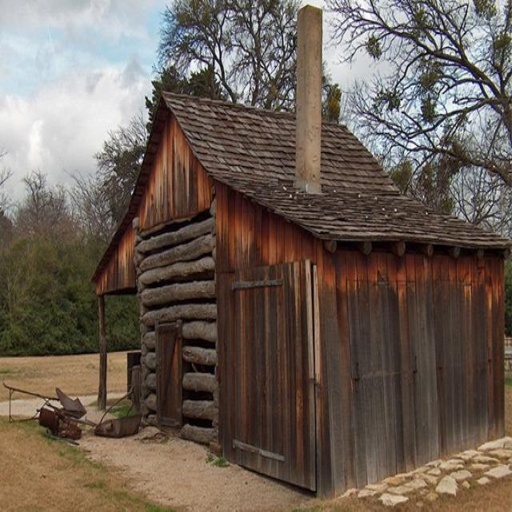}} 
\end{minipage}}	

\subfigure{ \label{cat}
\begin{minipage}[c]{\linewidth}\centering
\parbox[t]{0.15\linewidth}{
\includegraphics*[width=1\linewidth]{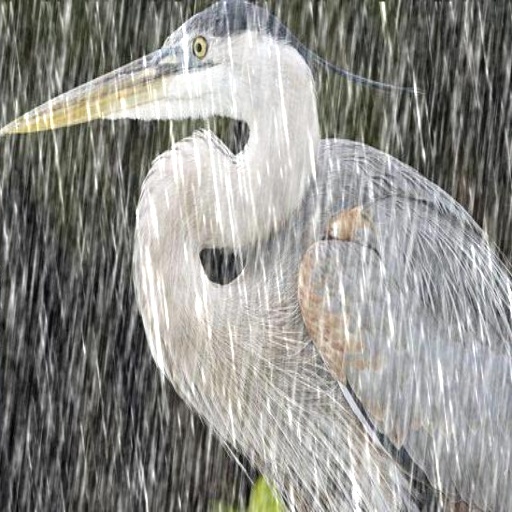}} 	
\parbox[t]{0.15\linewidth}{
\includegraphics*[width=1\linewidth]{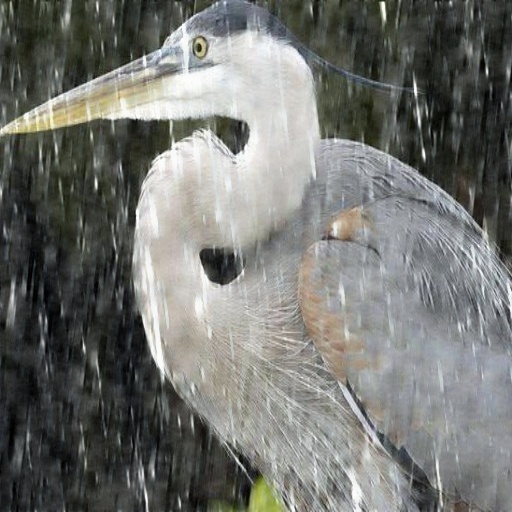}}  	
\parbox[t]{0.15\linewidth}{
\includegraphics*[width=1\linewidth]{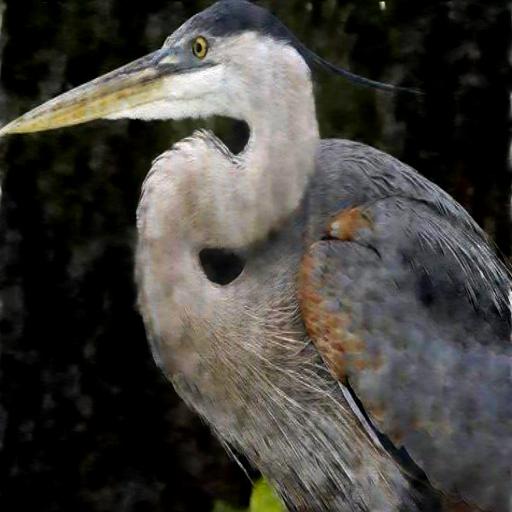}} 	
\parbox[t]{0.15\linewidth}{
\includegraphics*[width=1\linewidth]{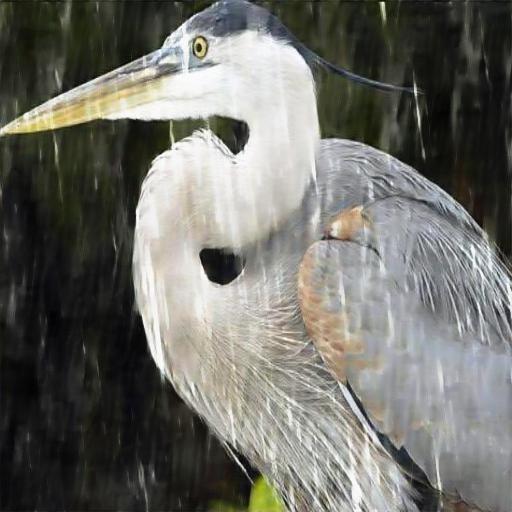}} 
\parbox[t]{0.15\linewidth}{
\includegraphics*[width=1\linewidth]{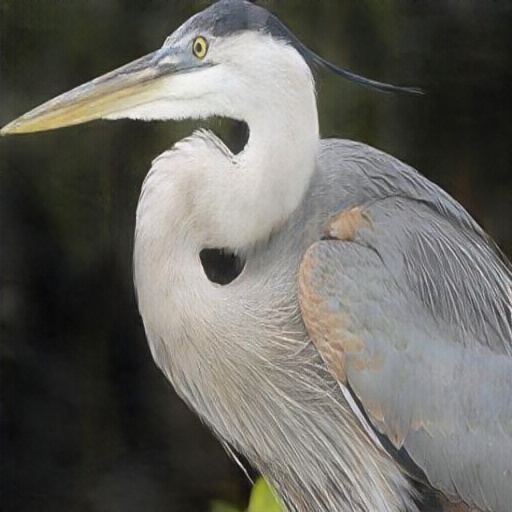}} 
\parbox[t]{0.15\linewidth}{
\includegraphics*[width=1\linewidth]{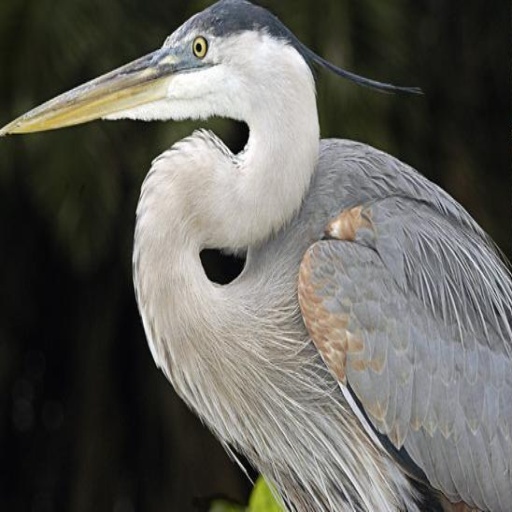}} 
\end{minipage}}

\parbox[center]{0.15\linewidth}{\centering \small(a) Input}
\parbox[center]{0.15\linewidth}{\centering \small(b) DDN}
\parbox[center]{0.15\linewidth}{\centering \small(c) JORDER}
\parbox[center]{0.15\linewidth}{\centering \small(d) DID-MDN}
\parbox[center]{0.15\linewidth}{\centering \small(e) DTDN}
\parbox[center]{0.15\linewidth}{\centering \small(f) Ground truth}

\caption{Results of different methods on sample images from $CRain-test$.} \label{different_methods_syn}
\end{figure*}

Quantitative and qualitative performances of different methods are compared on three testing sets: $Test1$, $Test2$ and $CRain\textrm{-}test$. In these three testing sets, $Test1$ contains simple rain streaks with single types and single directions; $Test2$ contains relatively complex rain streaks with single types and multiple directions; And $CRain\textrm{-}test$ includes more complex rain steaks with multiple types and multiple directions. We make use of these three datasets with different types and different directions of rain streaks to compare the ability of different methods to de-rain. Quantitative results of different methods are tabulated in Table~\ref{tab2}, Table~\ref{tab3} and Table~\ref{tab4}, respectively, and the results based on $CRain\textrm{-}test$ are shown in Figure~\ref{different_methods_syn}.

We train RESCAN~\cite{li2018recurrent} and JORDER~\cite{yang2017deep} on the enriched dataset ($CRain$) and train other models on the original dataset. Table~\ref{tab2}, Table~\ref{tab3} and Table~\ref{tab4} show that those models trained on the original dataset obtain unsatisfactory results in complex ($CRain\textrm{-}test$) and mid-complex ($Test\textrm{-}2$) rain conditions, although some models have commendable results in simple rain conditions. The reason why those models obtain low metrics in complex rain conditions from Figure~\ref{different_methods_syn}(b) and (d) lies in that those models cannot resist the interference of complex rain streaks during recovering the details of rainy images.

However, although we train RESCAN and JORDER on the enriched dataset, their results still decrease much when facing complex rain conditions. Based on Figure~\ref{different_methods_syn}(c), we find that JORDER damages color and textures when removing rain streaks. Generally, this is due to difficult optimization on complex datasets. Compared with other models, our proposed model achieves the best results.

\begin{table*}[h]
\centering
\newcommand{\tabincell}[2]{\begin{tabular}{@{}#1@{}}#2\end{tabular}}  
\caption{Quantitative results of different methods on $Test1$ in terms of PSNR, SSIM and UQI.}\label{tab2}
\begin{tabular}{|c|c|c|c|c|c|c|c|c|}
\hline
 & Input &  \tabincell{c}{CNN~\cite{fu2017clearing}} & \tabincell{c}{DDN~\cite{fu2017removing}} & \tabincell{c}{ID-CGAN~\cite{zhang2017image}} & \tabincell{c}{JORDER~\cite{yang2017deep}} & \tabincell{c}{DID-MDN~\cite{zhang2018density}} &RESCAN~\cite{li2018recurrent} & DTDN\\
\hline
PSNR &  22.11 &	22.06	&  26.46 & 22.05 & 25.44 & 27.98 & \textbf{30.77}& 29.01 \\
\hline
SSIM &	0.7771 & 0.8409	& 0.8731 & 0.8364 & 0.8691  & 0.9159 & \textbf{0.9296} & 0.9234 \\
\hline
UQI &	0.8548 & 0.8561	&  0.9210  & 0.8742 &   0.9256  & 0.9304 & \textbf{0.9454} & 0.9432\\
\hline
\end{tabular}
\end{table*}

\begin{table*}[h]
\centering
\newcommand{\tabincell}[2]{\begin{tabular}{@{}#1@{}}#2\end{tabular}}  
\caption{Quantitative results of different methods on $Test2$ in terms of PSNR, SSIM and UQI.}\label{tab3}
\begin{tabular}{|c|c|c|c|c|c|c|c|c|}
\hline
 & Input &  \tabincell{c}{CNN~\cite{fu2017clearing}} & \tabincell{c}{DDN~\cite{fu2017removing}} & \tabincell{c}{ID-CGAN~\cite{zhang2017image}} & \tabincell{c}{JORDER~\cite{yang2017deep}} & \tabincell{c}{DID-MDN~\cite{zhang2018density}} & RESCAN~\cite{li2018recurrent}  & DTDN\\
\hline
PSNR &  19.40 &	19.77	&  24.73 & 20.45 & 23.68 & 26.14 & 27.97 & \textbf{28.24} \\
\hline
SSIM &	0.7232 & 0.8041	& 0.8450 & 0.7926 & 0.8266 & 0.8855 & 0.8982 & \textbf{0.9097} \\
\hline
UQI &	0.8194 & 0.8211	&  0.9061  & 0.8513 & 0.8976 & 0.9203 &0.9294 &\textbf{0.9352}\\
\hline
\end{tabular}
\end{table*}

\begin{table*}[!h]
\centering
\newcommand{\tabincell}[2]{\begin{tabular}{@{}#1@{}}#2\end{tabular}}  
\caption{Quantitative results of different methods on $CRain-test$ in terms of PSNR, SSIM and UQI.}\label{tab4}
\begin{tabular}{|c|c|c|c|c|c|c|c|c|}
\hline
 & Input &  \tabincell{c}{CNN~\cite{fu2017clearing}} & \tabincell{c}{DDN~\cite{fu2017removing}} & \tabincell{c}{ID-CGAN~\cite{zhang2017image}} & \tabincell{c}{JORDER~\cite{yang2017deep}} & \tabincell{c}{DID-MDN~\cite{zhang2018density}} & RESCAN~\cite{li2018recurrent} &DTDN\\
\hline
PSNR &  17.30 &	17.9842	&  21.82 & 18.44 & 24.94 & 21.24 & 26.85 & \textbf{28.37}\\
\hline
SSIM &	0.6446 & 0.7356	& 0.7929 & 0.7181 & 0.8641 & 0.7962 & 0.8950 & \textbf{0.9183} \\
\hline
UQI &	0.8588 & 0.8564	&  0.9243  & 0.8744 &  0.9355 & 0.9055 & 0.9342 &\textbf{0.9538}\\
\hline
\end{tabular}
\end{table*}

\subsection{Results on Real-World (RW)}

\begin{figure*}[!h] 

\subfigure{ \label{cat}
\begin{minipage}[c]{\linewidth}\centering
\parbox[t]{0.15\linewidth}{
\includegraphics*[width=1\linewidth]{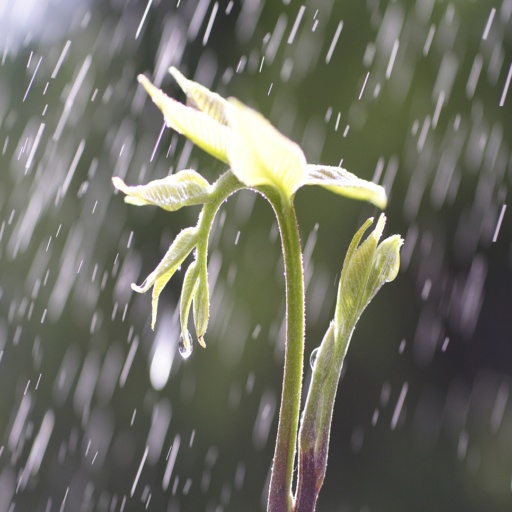}} 
\parbox[t]{0.15\linewidth}{
\includegraphics*[width=1\linewidth]{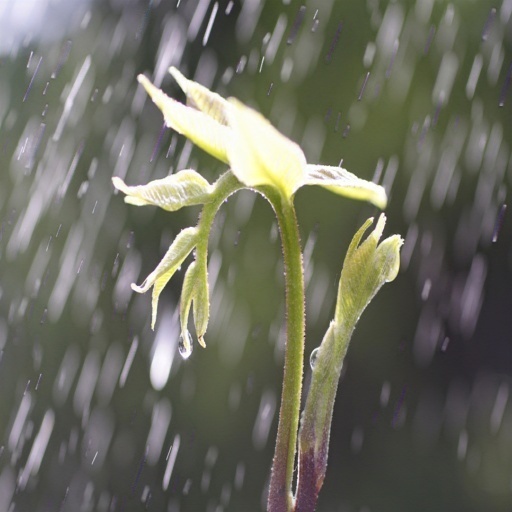}} 	
\parbox[t]{0.15\linewidth}{
\includegraphics*[width=1\linewidth]{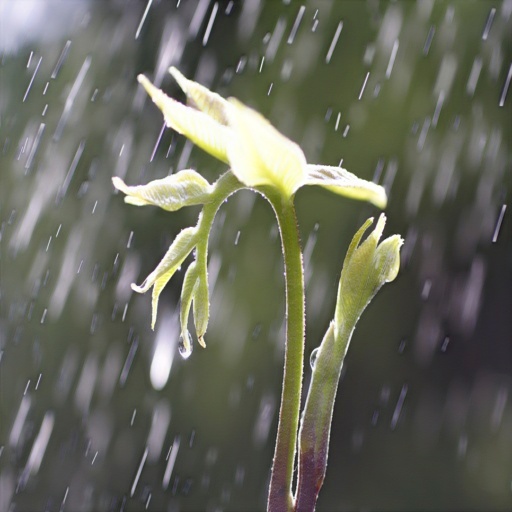}}  	
\parbox[t]{0.15\linewidth}{
\includegraphics*[width=1\linewidth]{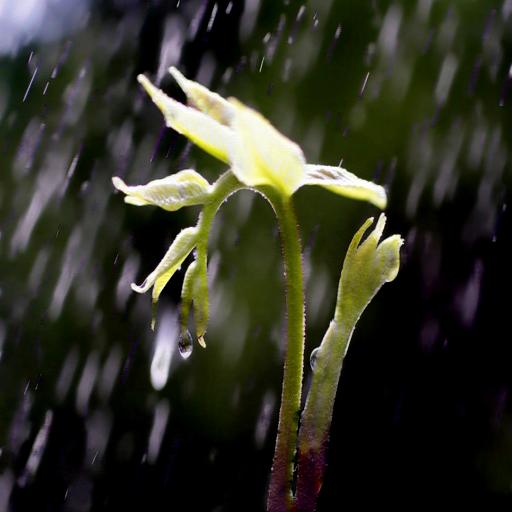}} 	
\parbox[t]{0.15\linewidth}{
\includegraphics*[width=1\linewidth]{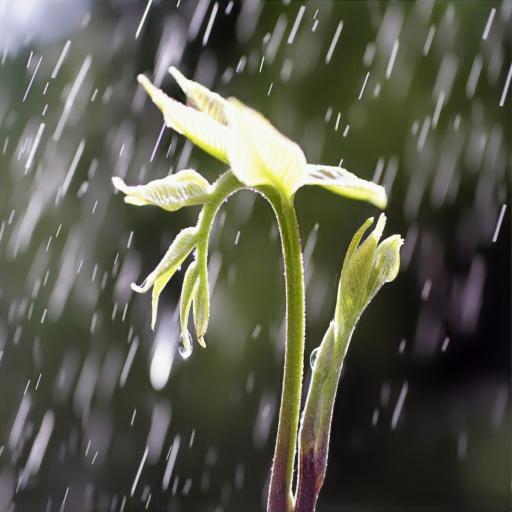}} 
\parbox[t]{0.15\linewidth}{
\includegraphics*[width=1\linewidth]{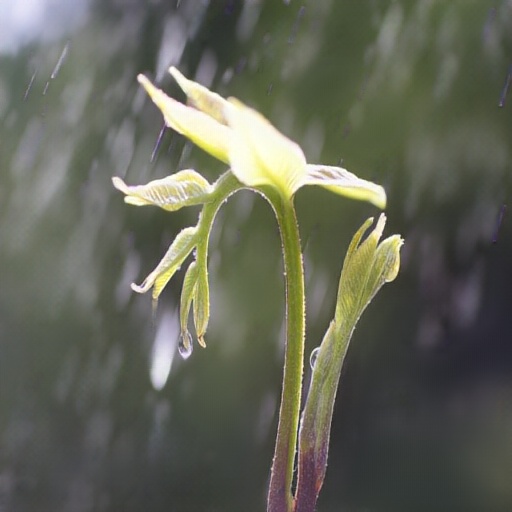}} 
\end{minipage}}	

\subfigure{ \label{cat}
\begin{minipage}[c]{\linewidth}\centering
\parbox[t]{0.15\linewidth}{
\includegraphics*[width=1\linewidth]{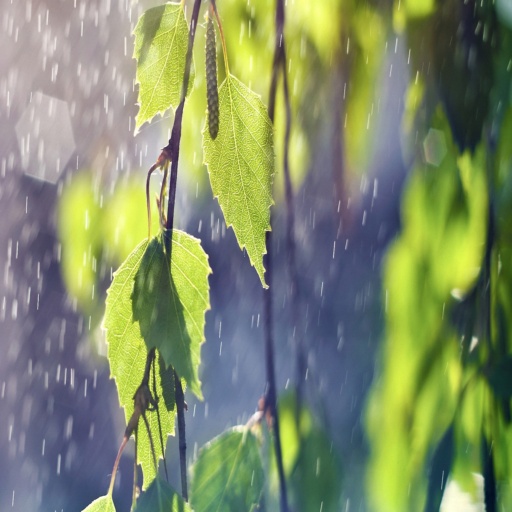}} 	
\parbox[t]{0.15\linewidth}{
\includegraphics*[width=1\linewidth]{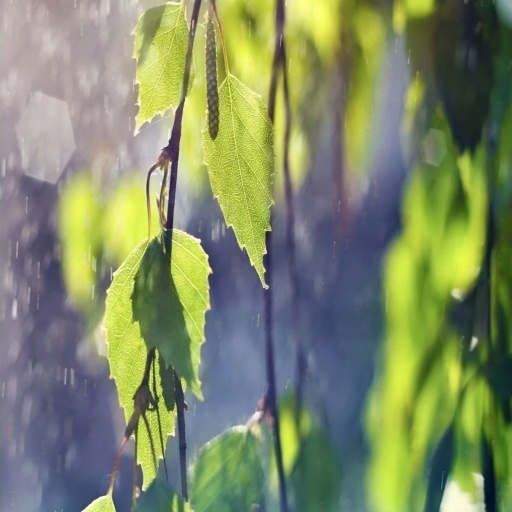}}  	
\parbox[t]{0.15\linewidth}{
\includegraphics*[width=1\linewidth]{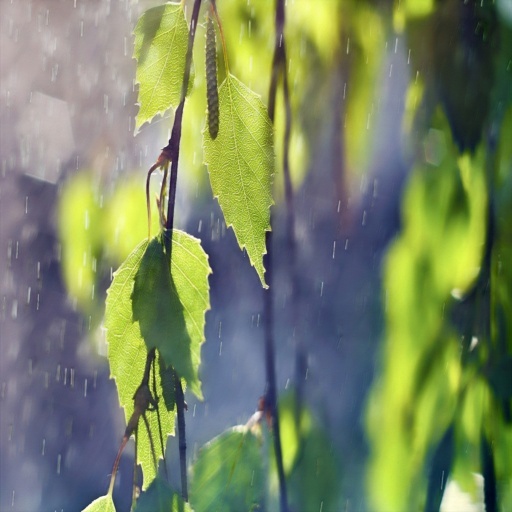}} 
\parbox[t]{0.15\linewidth}{
\includegraphics*[width=1\linewidth]{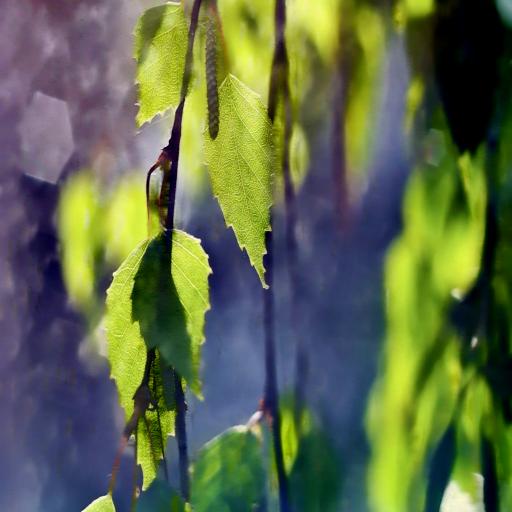}} 	
\parbox[t]{0.15\linewidth}{
\includegraphics*[width=1\linewidth]{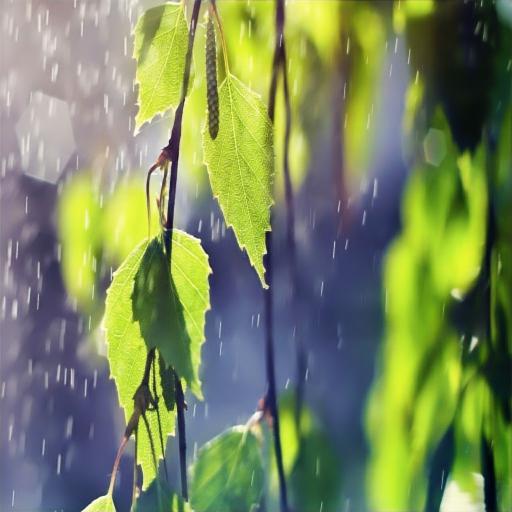}} 
\parbox[t]{0.15\linewidth}{
\includegraphics*[width=1\linewidth]{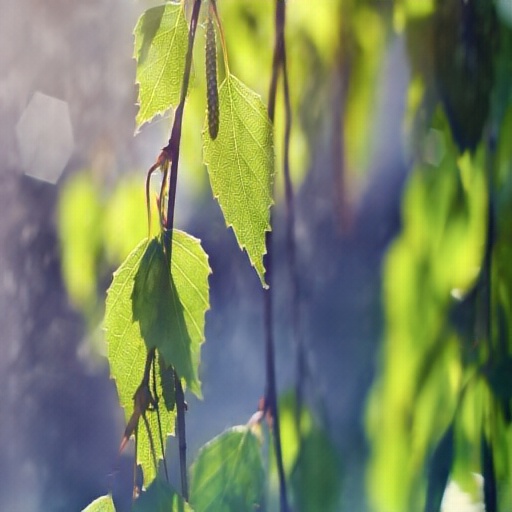}} 
\end{minipage}}	\vspace{0.1pt}


\subfigure{ \label{cat}
\begin{minipage}[c]{\linewidth}\centering
\parbox[t]{0.15\linewidth}{
\includegraphics*[width=1\linewidth]{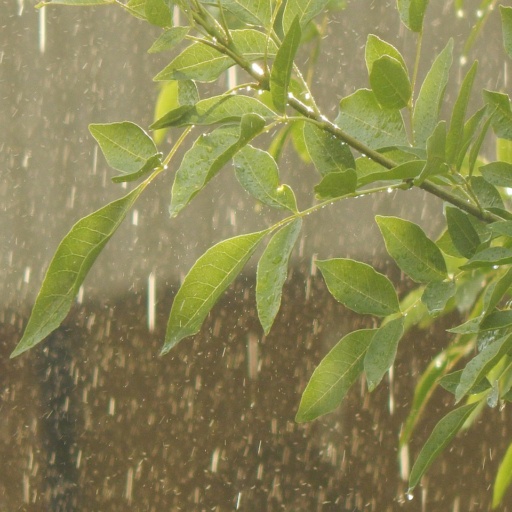}} 	
\parbox[t]{0.15\linewidth}{
\includegraphics*[width=1\linewidth]{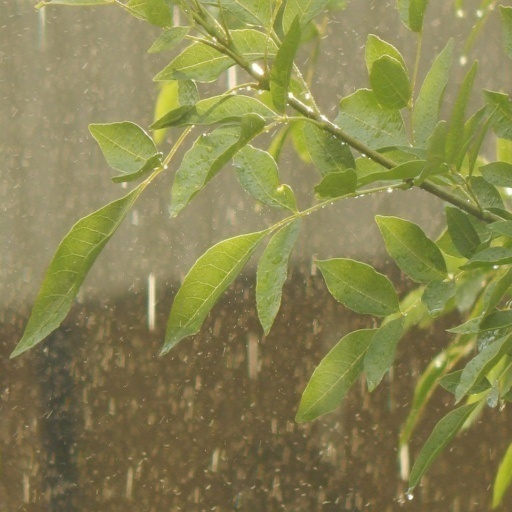}}  	
\parbox[t]{0.15\linewidth}{
\includegraphics*[width=1\linewidth]{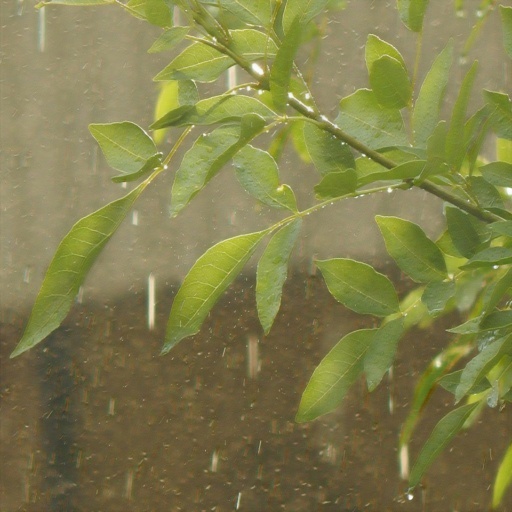}} 
\parbox[t]{0.15\linewidth}{
\includegraphics*[width=1\linewidth]{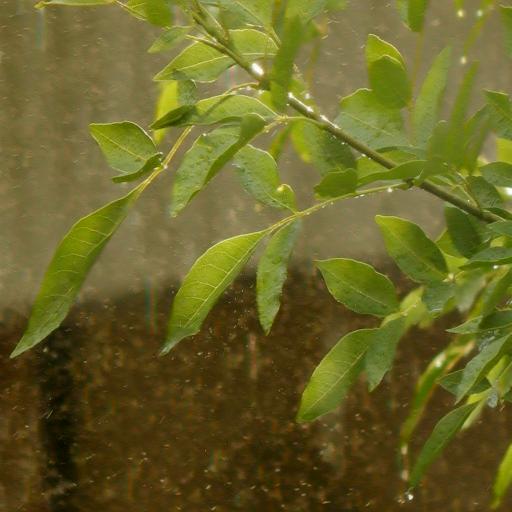}} 	
\parbox[t]{0.15\linewidth}{
\includegraphics*[width=1\linewidth]{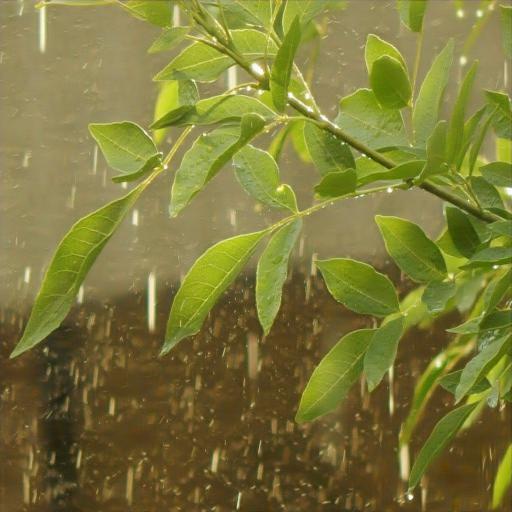}} 
\parbox[t]{0.15\linewidth}{
\includegraphics*[width=1\linewidth]{different_methods_real/ours/6}} 
\end{minipage}}	\vspace{0.1pt}

\subfigure{ \label{cat}
\begin{minipage}[c]{\linewidth}\centering
\parbox[t]{0.15\linewidth}{
\includegraphics*[width=1\linewidth]{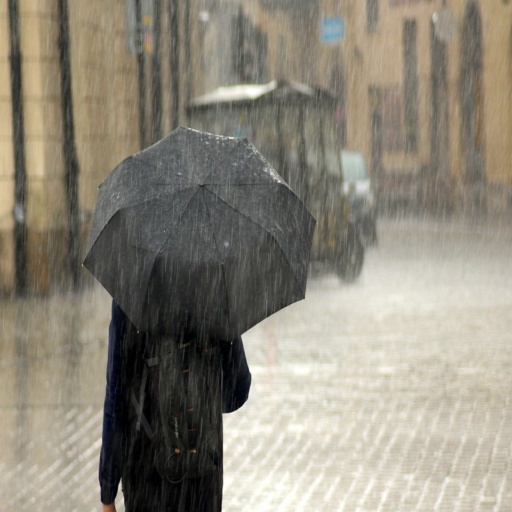}} 	
\parbox[t]{0.15\linewidth}{
\includegraphics*[width=1\linewidth]{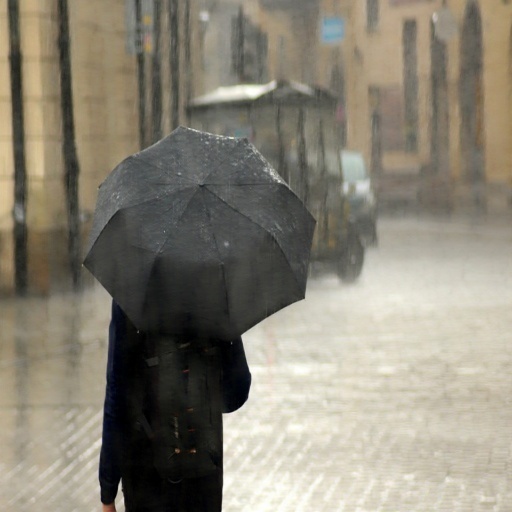}}  	
\parbox[t]{0.15\linewidth}{
\includegraphics*[width=1\linewidth]{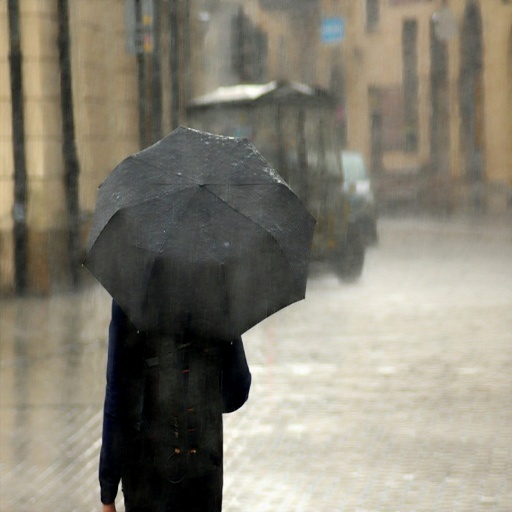}} 
\parbox[t]{0.15\linewidth}{
\includegraphics*[width=1\linewidth]{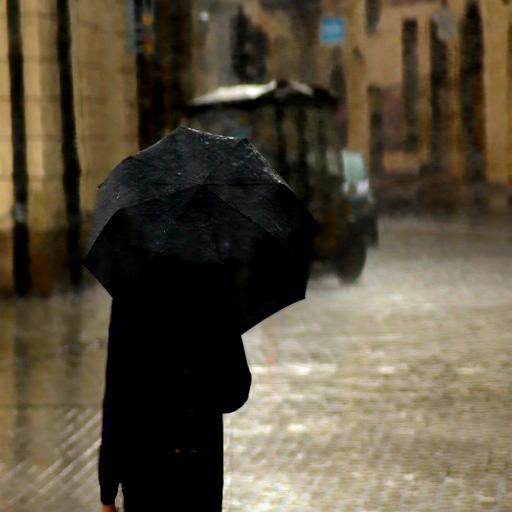}} 	
\parbox[t]{0.15\linewidth}{
\includegraphics*[width=1\linewidth]{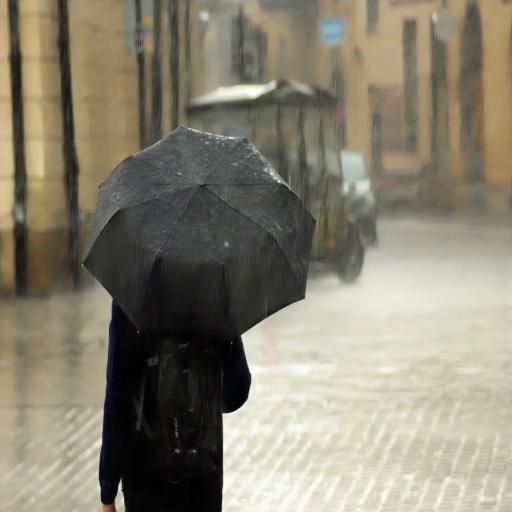}} 
\parbox[t]{0.15\linewidth}{
\includegraphics*[width=1\linewidth]{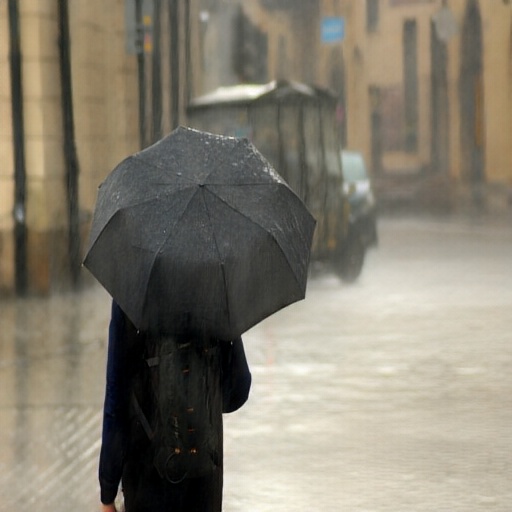}} 
\end{minipage}}

\parbox[center]{0.15\linewidth}{\centering \small(a) Input}
\parbox[center]{0.15\linewidth}{\centering \small(b) RESCAN}
\parbox[center]{0.15\linewidth}{\centering \small(c) DDN}
\parbox[center]{0.15\linewidth}{\centering \small(d) JORDER}
\parbox[center]{0.15\linewidth}{\centering \small(e) DID-MDN}
\parbox[center]{0.15\linewidth}{\centering \small(f) DTDN}

\caption{Results of different methods on samples from our real world dataset.} \label{different_methods_real}
\end{figure*}

Figure~\ref{different_methods_real} shows experimental results for real images. We select some images of typical rain conditions to compare different methods. The images in the first and second row have complex rain streaks caused by DOF, The images in the third row have complex rainy streaks caused by natural factors and the images in the last row have normal rain streaks.

JORDER~\cite{yang2017deep} does have a good performance in removing rain streaks but still damages some textures and details. Although RESCAN~\cite{li2018recurrent} has the highest metric values in $Test1$, but its performance in real images is unsatisfactory. As observed, our method significantly outperforms the others, not only removing the majority of rain streaks but also retaining the texture and details information.

\section{Conclusion}

This paper proposes an end-to-end network DTDN and a specific training algorithm for single image de-raining. DTDN is composed of a GAN mainly for removing rain streaks and a CNN mainly for recovering the details from original images. The training algorithm increases the possibility of obtaining a well-refined de-raining network. We further propose a technique to enrich existing datasets for better modeling. 

Experimental results show that DTDN is more effective for single image de-raining, compared to other state-of-the-art methods.
The coupled system of GAN and CNN is a framework of solving such problems with two mutually exclusive modeling tasks. In future research, we will apply the proposed model to image denoising, image super-resolution, etc.
%
\bibliographystyle{ACM-Reference-Format}
\balance
\bibliography{acmart}

\end{document}